\documentclass[conference,compsoc]{IEEEtran}
\usepackage[latin1]{inputenc}
\usepackage[T1]{fontenc}

\usepackage{verbatim}
\usepackage{amsmath}
\usepackage[linesnumbered,ruled,vlined]{algorithm2e}

\SetCommentSty{mycommfont}
\usepackage{amssymb}

\usepackage{graphicx}
\usepackage{epstopdf}
\usepackage{subfigure}
\usepackage{array}
\usepackage{booktabs}
\usepackage{multirow}
\usepackage{caption}
\graphicspath{{images/}}

\usepackage{listings}

\usepackage{enumitem}

\usepackage{xcolor}
\usepackage{colortbl} %
\definecolor{jade}{rgb}{0.0, 0.66, 0.42}
\definecolor{carolinablue}{rgb}{0.6, 0.73, 0.89}
\definecolor{dkgreen}{rgb}{0,0.6,0}
\definecolor{dkblue}{rgb}{0,0.4,0.5}
\definecolor{gray}{rgb}{0.5,0.5,0.5}
\definecolor{mauve}{rgb}{0.58,0,0.82}

\usepackage[final]{changes}

\setaddedmarkup{\color{blue}{#1}}
\setdeletedmarkup{\color{orange}{\sout{#1}}}

\usepackage{url}
\definecolor{mypink1}{RGB}{237, 2, 140}
\usepackage[breaklinks,colorlinks,bookmarks=false]{hyperref}
\hypersetup{citecolor=blue,linkcolor=red,urlcolor=mypink1}

\usepackage{flushend}

\usepackage{framed}

\usepackage{pifont}
\usepackage[misc]{ifsym}
\usepackage{bbding}
\usepackage{fancyhdr} %
\usepackage{bibspacing}
\usepackage{threeparttable}
\usepackage{booktabs} %
\usepackage{xspace}

\newcommand{\tabincell}[2]{\begin{tabular}{@{}#1@{}}#2\end{tabular}}

\newcommand{\sysname}{\textsc{Callee}\xspace}

\begin{document}

\date{}

\title{\large \sysname: Recovering Call Graphs for Binaries with Transfer and Contrastive Learning}

\author{\IEEEauthorblockN{Wenyu Zhu$^{\dagger\ddagger}$, Zhiyao Feng$^{\dagger\ddagger}$, Zihan Zhang$^{\dagger\ddagger}$, Jianjun Chen$^{\dagger\S}$, Zhijian Ou$^\dagger$, Min Yang$^\star$, Chao Zhang$^{\dagger\ddagger\S\ast}$}
\IEEEauthorblockA{$\dagger$ \textit{Tsinghua University, Beijing, China} $^\ddagger$ \textit{BNRist} $^\S$ \textit{Zhongguancun Laboratory}}
\IEEEauthorblockA{$^\star$ \textit{Fudan University, Shanghai, China}
\IEEEauthorblockA{$^\ast$ \textit{Corresponding author}}
} 
}

\maketitle

\begin{abstract}
Recovering binary programs' call graphs is crucial for inter-procedural analysis tasks and applications based on them.
One of the core challenges is recognizing targets of indirect calls (i.e., indirect callees).
Existing solutions all have high false positives and negatives, making call graphs inaccurate.
In this paper, we propose a new solution \sysname combining transfer learning and contrastive learning.
The key insight is that, 
deep neural networks (DNNs) can automatically identify patterns concerning indirect calls.
Inspired by the advances in question-answering applications, 
we utilize \textit{contrastive} learning to answer the callsite-callee question.
However, one of the toughest challenges is 
that DNNs need large datasets to achieve high performance,
while collecting large-scale indirect-call ground truths can be computational-expensive.
Therefore, we leverage \textit{transfer} learning to pre-train DNNs with easy-to-collect direct calls 
and further fine-tune DNNs for indirect-calls.
We evaluate \sysname on several groups of targets,
and results show that our solution could match callsites to callees with an \textit{F1-Measure} of 94.6\%, much better than state-of-the-art solutions.
Further, we apply \sysname to two applications -- binary code similarity detection and hybrid fuzzing,
and found it could greatly improve their performance.

\end{abstract}

\vspace{-0.1cm}
\section{Introduction} \label{introduction}
\vspace{-0.1cm}
Indirect calls (\textit{icalls} for short) allow programs to determine the choice of functions to call (i.e., callees) until runtime, enabling programmers to realize dynamic features,
and thus are commonly used in object-oriented programming as well as some large-scale programs such as the Linux kernel.  
Meanwhile, icalls play an important role in program analysis and related tasks. 
One can complement Call Graphs (CGs) of programs by recognizing targets of indirect calls
(\textit{icallees} for short), 
and many tasks can benefit from precise CGs such as inter-procedural data-flow analysis~\cite{shi2018pinpoint}, 
binary code similarity detection~\cite{xu2017neural}, and even test case generation for fuzzing~\cite{ndss2020hfl}.
For example, SelectiveTaint\cite{chen2021selectivetaint} 
relies on CG reconstruction for taint analysis, 
$\alpha$Diff~\cite{liu2018alphadiff} and DeepBinDiff~\cite{duan2020deepbindiff} perform binary diffing with CG features, 
and TEEREX\cite{cloosters2020teerex} requires precise CGs to perform symbolic execution. 
Conversely, imprecise icallee analysis will lead to obstacles in many applications,
such as false positives in bug detection~\cite{jana2016automatically,kang2016apex,xu2018precise}
and path explosion in symbolic execution~\cite{s2eSystem,angr}.

In practice, it is common to utilize static analysis to infer icallees,
because dynamic techniques can miss many legitimate callees due to poor code coverage
\deleted{, though they have no false positives}.
Given target programs with or without source code, 
applicable static analysis solutions are different.
When the source code is available, 
points-to analysis~\cite{sui2016svf,sui2018value} and type-based analysis~\cite{tice2014enforcing,lu2019does} are the most common methods. 
{\em \deleted{If only binaries are given,} Otherwise statically
determining icallees is much more challenging, }
since much information (e.g., type) is missing \added{in binaries}.

Existing binary-level solutions in general apply an approximation algorithm to recognize icallees.
For instance, 
binary analysis tools that are widely used in practice (e.g., IDA Pro~\cite{IDAPro}, Angr~\cite{angr}, GHIDRA~\cite{Ghidra}) 
and PathArmor~\cite{van2015patharmor}
identify icallees by constant propagation, and can only resolve very few targets.
On the other hand, CCFIR~\cite{ccfir} adopts
the address-taken policy and treats all address-taken functions as potential icallees, 
thus having high false positives. 
$\tau$CFI~\cite{muntean2018taucfi},
TypeArmor~\cite{van2016typearmor} and its refinement~\cite{lin2021improvetypearmor} 
reduce icallees to reduce false positives
by first recovering function prototypes 
and then performing type-based matching, but have low guarantees of correctness. 
The state-of-the-art solution BPA~\cite{kim2021icall} performs a delicate pointer analysis based on a block memory model and a special intermediate representation language (with only support for x86) to infer icallees, 
but the prototype did not support C++ binaries and still has relatively low precision.
A better solution to recognize icallees in binaries is therefore demanded.

In this paper, we propose a deep-learning solution \sysname to recognize icallees at the binary level.
Given an indirect callsite (\textit{icallsite} for short), 
\sysname will answer which callees could be its potential targets.
{\em The key insight is that, 
with sufficient data, \deleted{deep neural networks (DNNs)}\added{DNNs} can automatically identify patterns concerning icalls, which can be much more efficient than introducing approximation algorithms or heuristic rules to handle various cases.}
Specifically, \added{combining contrastive learning and transfer learning,} DNNs can learn to match callsites and callees by comprehending their contexts, i.e., instructions nearby callsites and of callees.

Contrastive learning 
\deleted{is a machine-learning technique that 
encourages similar inputs to have similar representations }
\added{aims to represent similar inputs with similar embeddings}
in the latent space, 
and has been proved effective in question-answering scenarios \cite{yu2014deep,wang2015long}.
\added{Thus} regarding a callsite as a question and a callee as its corresponding answer,
we build a contrastive-learning framework to match callsites with callees. 
Beforehand, we perform slicing to extract instructions for callsites and callees based on the calling convention, and \added{embed} the generated slices\deleted{are transformed into vectors} 
by adjusting a popular representation learning technique doc2vec~\cite{le2014distributed} 
to the assembly language.
In addition, we propose a new symbolization policy to symbolize assembly tokens to 
improve the model performance and meanwhile handle the out-of-vocabulary (OOV) problem.

\added{Moreover}\deleted{However}, DNNs need large datasets to achieve high performance,
while collecting icall ground truths requires \added{computational-expensive} dynamic analyses
\deleted{, which is computational-expensive}.
Whereas direct calls (\textit{dcalls} for short) can be easily obtained with static analyses.
Thus it would be exceedingly beneficial if we can train DNNs for icalls with\deleted{the help of} dcalls, \added{i.e., transfer learning, which reuses}
\deleted{Transfer learning  is a machine learning technique reusing} a pre-trained model for one task as the starting point for a model on another task. 
It has been proved efficient to transfer knowledge between languages, images and voices~\cite{zhuang2020transferlearningsurvey},
and recently in program analysis~\cite{li2022plato}. 
Considering that dcalls and icalls share similar calling conventions,
it is possible to transfer knowledge learned from dcalls to icalls.
Therefore, we leverage transfer learning to train DNNs for icalls based on abundant dcalls.
Specifically, we utilize contrastive learning to answer the callsite-callee question for both dcalls and icalls, while the icall DNN is initialized with a pre-trained dcall DNN.

We have implemented a prototype of \sysname
and evaluated it on targets that have abundant icalls such as the Linux kernel and the Firefox browser~\cite{firefox}.
The evaluation results show that \sysname could match callsites to callees with an \textit{F1-Measure} (\textit{F1}) of 94.6\%,
, recall of 90.9\%, and precision of 97.3\%,
outperforming BPA~\cite{kim2021icall}, 
TypeArmor~\cite{van2016typearmor} 
as well as real-world binary analysis tools \added{such as} IDA Pro~\cite{IDAPro}
\deleted{, Angr~\cite{angr} and GHIDRA~\cite{Ghidra}}.

Further, we have demonstrated that \sysname can benefit down-stream applications based on call graphs.
Firstly, we applied \sysname to binary code similarity detection
and greatly improved the state-of-the-art solution DeepBinDiff
with an average increase of 4.6\% \textit{F1} in cross-version binary diffing 
and 13.7\% in cross-optimization binary diffing.
Moreover, \sysname is applied to the widely-used hybrid fuzzing solution Driller~\cite{stephens2016driller}.
In three 24-hour fuzzing campaigns, it can help the fuzzer find 50\% more paths\deleted{and unique crashes} 
on average in all 8 CGC~\cite{darpa2014cgc} challenges that have icalls.

Additionally, we have made an attempt to interpret the neural network with a case study
where \sysname surpassed other solutions.
It showed that the proposed model can well capture semantic features of tokens in assembly instructions,
and tokens related to arguments and return values contribute the most to icallee recognition,
which is consistent with the domain knowledge of binary analysis.

In summary, we make the following contributions:
\begin{itemize}[leftmargin=.35cm,noitemsep,topsep=0pt]
    \item 	We present the first transfer- and contrastive-learning approach \sysname integrated with expert knowledge to recognize icallees and recover call graphs for binaries.
    \item   We propose a new symbolization method for machine-learning solutions on assembly language, which can preserve data-flow information of assembly contexts and meanwhile does not introduce the OOV problem.
    \item   We have collected the largest set of callsite-callee training data.
    The dataset and nerual model are available at \href{https://github.com/vul337/Callee}{https://github.com/vul337/Callee}.
    \item 	We evaluate \sysname with real-world programs
    and demonstrate that it outperforms state-of-the-art solutions
    on the callsite-callee matching task.
    \item   We demonstrate that \sysname is highly effective at promoting tasks based on CGs, e.g., binary code similarity detection or hybrid fuzzing. 
\end{itemize}

\vspace{-0.1cm}
\section{Background and Related Work} \label{Background}

\vspace{-0.1cm}
\subsection{Transfer Learning}
\vspace{-0.1cm}

Given a source domain $D_S=\{X_S, f_S(X,\theta_S)\}$ and learning task $T_S$, 
a target domain $D_T=\{X_T, f_T(X,\theta_T)\}$ and learning task $T_T$, transfer learning aims to help improve the learning of the target predictive function $f_T$ in $D_T$ using the knowledge in $D_S$ and $T_S$, 
where $D_S\neq D_T$, or $T_S\neq T_T$.
In general, one of the most common methods to perform transfer learning is to initialize $f_T$
with parameters of the pre-trained $f_S$, i.e. using $\theta_S$ as the initial value of $\theta_T$.

In fields of Natural Language Processing (NLP), 
transfer-learning techniques \cite{gururangan-etal-2020-dont} have been
proposed to transfer knowledge between two languages (e.g., English and Nepali).
Recently, PLATO \cite{li2022plato} proposed a cross-lingual transfer-learning framework for
statistical type inference in source code. 
And StateFormer~\cite{pei2021stateformer} utilized a pretrain-finetune
architecture to recover function type information from assembly code,
shedding light on applications of transfer learning on program analysis.

\vspace{-0.1cm}
\subsection{Contrastive Learning}
\vspace{-0.1cm}
Contrastive learning aims to teach a neural model to pull together 
the representations of matching samples in a latent space, 
and meanwhile separate non-matching ones.
The most common method is through a Siamese network~\cite{bromley1994signature},
which is a structure with two parallel networks to extract feature vectors of two input samples,
and calculate the distance with another neural network or pre-defined norms. 
At first, the Siamese network was proposed to compare the similarity of two inputs. 
It consists of two identical networks with identical structures and weights.
Distance between the feature vectors of inputs will be calculated and used as the similarity/difference score.
Previous studies such as $\alpha$Diff~\cite{liu2018alphadiff} and NMT~\cite{zuo2019neural} have shown that the Siamese network could be utilized to extract fine-grained semantic features of binary code,
even if the code is from cross-version or cross-architecture binaries.

\begin{figure}[b]
    \centering
    \includegraphics[width = .62\linewidth]{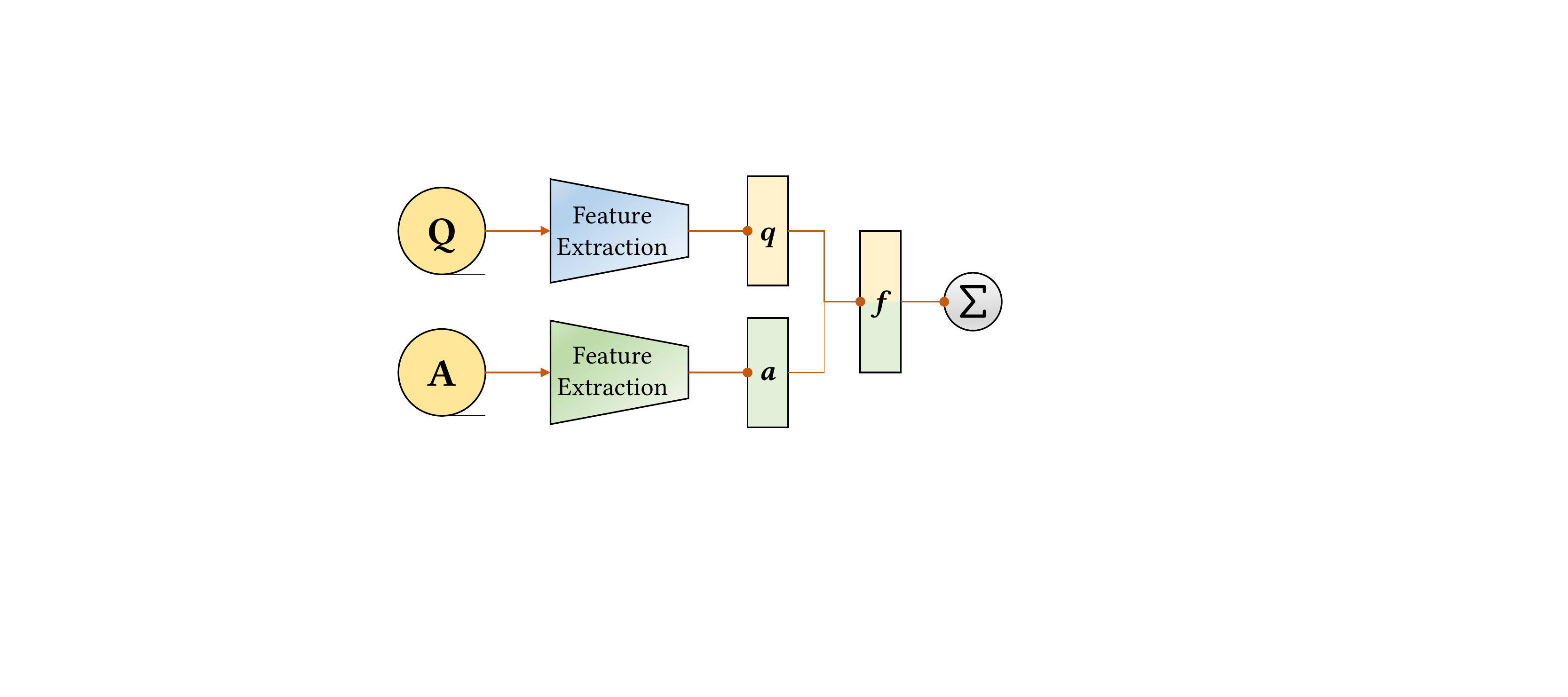}
    \caption{Illustration of the Siamese network}
    \label{fig:siamese}
\vspace{-0.5cm}
\end{figure}

Recently, another type of Siamese network is introduced to address more complicated problems. 
The new structure, also called a pseudo-Siamese network,
allows two networks to be different or not to share weights 
to adapt to application scenarios which require different categories of inputs. 
As shown in Figure~\ref{fig:siamese}, 
in the question-answering scenario, 
two different networks can be utilized to extract features of a question (\textbf{Q}) and an answer (\textbf{A}) respectively.
To calculate the similarity/difference, the extracted feature vectors
\textit{\textbf{q}} and \textit{\textbf{a}} could be
concatenated together as a feature vector \textit{\textbf{f}}, 
which will be further fed into a following classifier network \textbf{$\Sigma$}.
The classifier will output a score indicating how much \textbf{Q} and \textbf{A} matches. 
This structure could be trained to match questions with answers, as shown in~\cite{minaee2017automatic, yu2017QA1,zhao2019simpleQA}.

\vspace{-0.1cm}
\subsection{Applications based on Call Graphs}
\vspace{-0.1cm}
Binary program analysis applications often have to track data flow between functions to comprehend the semantics of programs, and thus have to conduct inter-procedural program analysis
by traversing programs' Call Graphs (CGs)
which represent functions calling relationships 
to track information flow or capture the semantics.
Such applications include but are not limited to the followings.

\textbf{Binary Similarity Detection.}
BinDiff~\cite{BinDiff} matches functions based on their position or neighborhoods in CGs.
$\alpha$Diff~\cite{liu2018alphadiff} extracts inter-function and inter-module features based on CGs, and further calculates feature distances with a Siamese neural network.
DeepBinDiff~\cite{duan2020deepbindiff} utilizes CGs to construct inter-procedural control-flow graphs (ICFGs) and performs random walks on them to embed each basic block.

\textbf{Hybrid Fuzzing.}
Driller~\cite{stephens2016driller} leverage symbolic execution engines to solve inputs for program paths when the AFL \cite{AFL} fuzzer gets stuck, and SymQemu~\cite{poeplau2021symqemu} further proposes
a compilation-based symbolic execution policy to boost the speed of the symbolic executor.
However, they do not resolve icallees due to the path-explosion problem.
Thus by providing symbolic execution engines with a limited set of candidate targets,
we can ease the path-explosion problem and thereby enable hybrid fuzzers to resolve icallees to improve the code coverage.

Except for aforementioned applications, CGs are also vital in 
malware detection~\cite{hu2009SMIT},
bug detection~\cite{bai2020idea,shen2019neuex}
and many other scenarios~\cite{zhao2020patchscope,almakhdhub2020murai,xi2019deepintent}.

{\em Therefore the completeness and accuracy of CGs greatly affect the results of these applications. }
Otherwise, it may cause issues like false positives in bug detection,
path explosion in symbolic execution, etc.

\begin{figure*}[t]
    \centering
    \setlength{\abovecaptionskip}{0cm}
    \includegraphics[width=.72\textwidth]{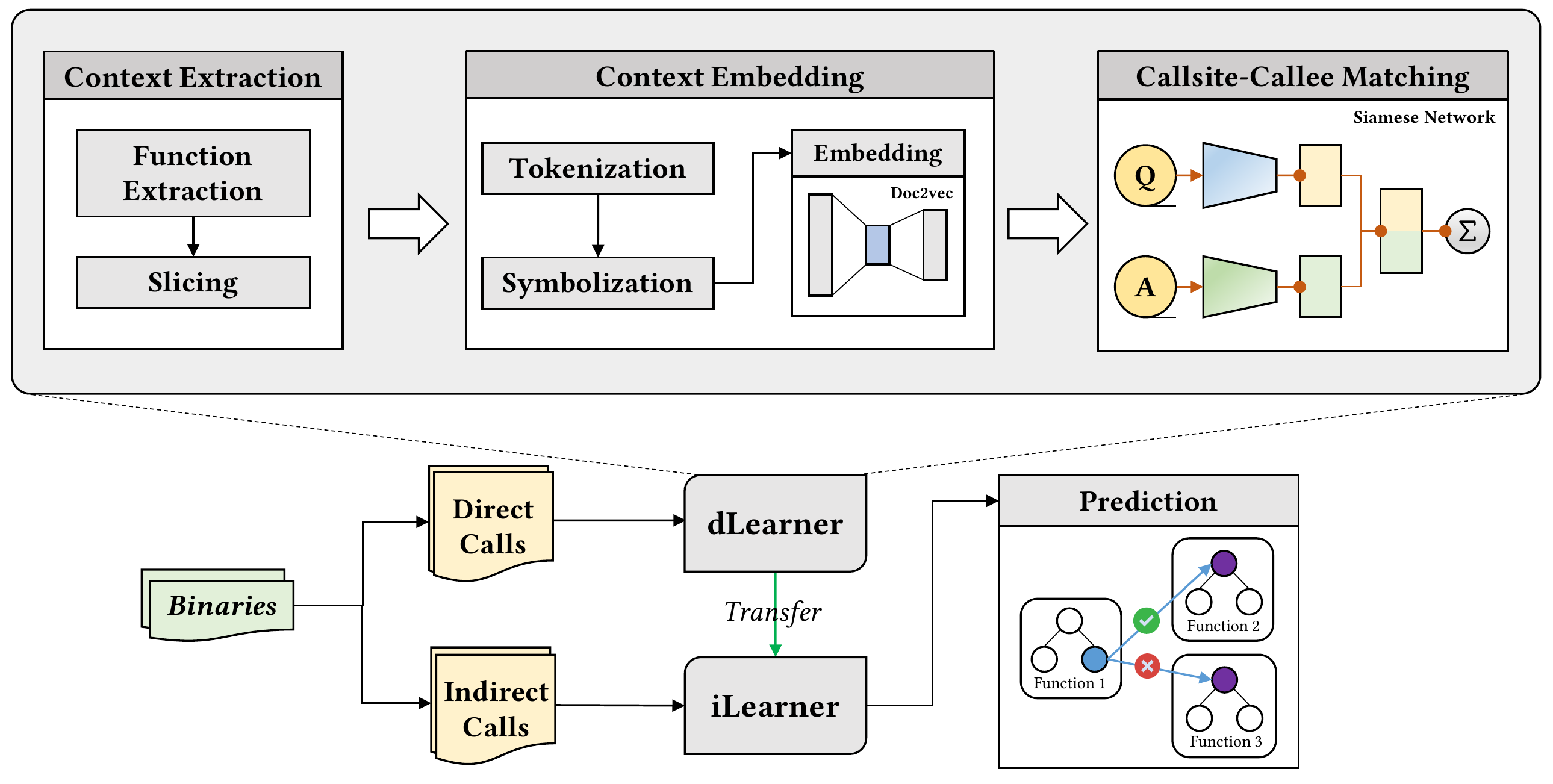}
    \caption{Overview of our solution \sysname.}
    \label{fig:overview}
    \vspace{-0.4cm}
\end{figure*}

\vspace{-0.1cm}
\subsection{Recognizing Indirect Callees in Binaries}
\vspace{-0.1cm}
At the core, constructing a complete and accurate CG requires to precisely recognize icallees.
Many solutions have been proposed to address this problem,
but few can recognize icallees for binaries.

\textbf{Type-based Analysis.}
Identifying icallees in binary programs in general requires type recovery analysis~\cite{lee2011tie} which is error-prone, as shown in $\tau$CFI~\cite{muntean2018taucfi}, TypeArmor~\cite{van2016typearmor} and its refinement \cite{lin2021improvetypearmor}. 
Otherwise, a coarse-grained address-taken policy would be applied, as shown in CCFIR~\cite{ccfir}, in which arbitrary address-taken functions are marked as legitimate icallees, causing more false positives.

\textbf{Pointer Analysis.} 
SVF~\cite{sui2016svf} leverages Andersen's algorithm and constructs an inter-procedural static single assignment (SSA) form to capture def-use chains of both top-level and address-taken variables,
While whole-program analyses such as SVF and SUPA~\cite{sui2018value} have troubles on programs composed of separately compiled modules. 
K-Miner~\cite{gens2018kminer} splits kernel code based on system calls, 
and PeX~\cite{zhang2019pex} leverages the common programming paradigm used in kernel abstraction interfaces, but they have not scaled to user-mode binaries.
Some binary analysis tools such as BAP~\cite{brumley2011bap} and Angr~\cite{angr} leverage value-set analysis to resolve pointers, but face challenges on complex real-world programs.
BDA~\cite{zhang2019bda} proposes a path sampling algorithm to perform dependency analysis
while introducing huge runtime overhead, even more than dynamic testing on multiple targets from
the SPECINT2000 benchmark~\cite{peng2014xforce}, making it impractical. 
Recently, BPA \cite{kim2021icall} adds scalable pointer analysis support for binaries based on a special block memory model and intermediate representation (IR), 
while the prototype currently supports 32bit C programs only.

\vspace{-0.1cm}
\subsection{DNN-based Binary Analysis}
\vspace{-0.1cm}
Recent research has leveraged DNNs to solve many program analysis problems. 

\textbf{Function Recovery.} 
Shin et al.~\cite{shin2015recognizing} show that recurrent neural networks (RNNs) can identify functions in binaries precisely.
It converts each byte into a vector with one-hot encoding, 
and concatenates vectors of all bytes as the representation of functions. 
Then it trains an RNN and uses the softmax function to predict whether a byte begins (or ends) a function.
XDA~\cite{pei2020xda} improves the performance by applying a BERT~\cite{devlin2018bert} model.
EKLAVYA~\cite{chua2017neural} and StateFormer~\cite{pei2021stateformer} further recovers function signatures from assembly code. 
EKLAVYA embeds each instruction into a vector and concatenates them to represent functions, 
and predicts a type tuple for all the parameters of a function with an RNN.
StateFormer~\cite{pei2021stateformer} utilizes transfer learning with a transformer~\cite{vaswani2017transformer} model to learn type inference rules.
However, they both cannot recover the signature of a callsite, and thus cannot recognize icallees. 

\textbf{Value-set Analysis (VSA).} 
DEEPVSA~\cite{guo2019deepvsa} uses DNNs to facilitate VSA 
by learning semantics of instructions and capturing dependencies in contexts at the binary level, 
which can further assist alias analysis for crash diagnosis. 
But the application in resolving icallees needs further study.

\textbf{Binary Similarity Detection.}
$\alpha$Diff first utilizes a DNN to learns code features from raw bytes, then extracts inter-function and inter-module features and adopts a Siamese neural network to detect similarity between binaries.
BinaryAI~\cite{yu2020ordermatters} uses BERT to pre-train the binary code on several tasks and adopts convolutional neural network (CNN) to extract the order information of CFG's nodes.
NMT~\cite{zuo2019neural} proposes a DNN-based cross-lingual basic-block embedding model to measure the similarity of two blocks, which achieves cross-architecture similarity detection. 
By regarding instructions as words and basic blocks as sentences, they use word2vec~\cite{mikolov2013distributed} to embed instructions and use LSTM~\cite{hochreiter1997lstm} to embed basic-blocks. 
The state-of-the-art DeepBinDiff~\cite{duan2020deepbindiff} uses both the code semantics and the program-wide control-flow information to generate basic block embedding. 

To the best of our knowledge, {\em we are the first to use deep learning to comprehend contexts of call instructions and recognize icallees}, and utilize it to recover CGs for binaries with a high precision.

\vspace{-0.1cm}
\section{Overview} \label{Overview}
\vspace{-0.1cm}

Our goal is to design a callsite-callee matching system that can automatically recognize which callees are potential targets for a given callsite.
In this section, we describe the overview of our solution \sysname.

\textbf{Overall workflow.} 
As shown in Figure~\ref{fig:overview}, we first train a contrastive-learning framework \textbf{Learner} with dcalls (denoted by \textbf{dLearner}), and transfer the learned knowledge into a icall \textbf{Learner} (denoted by \textbf{iLearner}).
In detail, parameters of the \textbf{iLearner} are initialized with the
pre-trained \textbf{dLearner}.
The \textbf{iLearner} will further be trained with icalls 
and used to perform icallee prediction.
To build such a \textbf{Learner} framework, 
we employ three major modules, i.e., context extraction module, context embedding module, and callsite-callee matching module.
The key insight is that, 
{\em neural networks can learn to match callsites with callees by comprehending their contexts, i.e., instructions nearby callsites and of callees.}

\vspace{-0.1cm}
\subsection{Core Modules of the Learner}

\vspace{-0.1cm}
\subsubsection{Context Extraction}\quad
Contexts related to callsites and callees form the basis of decisions made by neural networks.
Therefore, given a binary program, we first need to extract proper contexts from the binary.
Full contexts, i.e., all instructions of a function, 
make it difficult to construct favorable embeddings of limited vector dimensions. 
Therefore, shrinking the contexts while keeping necessary information is critical.
We adopt inter-procedural slicing with expert knowledge to extract related contexts.

\vspace{-0.1cm}
\subsubsection{Context Embedding}\quad
Since neural networks require vectors as inputs, contexts of callsites and callees have to be represented in the form of vectors. 
Existing studies~\cite{liu2018alphadiff} have shown that NLP solutions are effective at binary analysis.
We thus utilize a popular NLP model doc2vec to embed program slices.
Moreover, we adjust the doc2vec model with domain knowledge, i.e., differences between assembly and natual languages.

\vspace{-0.1cm}
\subsubsection{Callsite-callee Matching}\quad 
Inspired by question-answering scenarios, 
\sysname regards a callsite as a question and a callee as its corresponding answer.
To compute the difference score of a callsite and a callee, \sysname adopts contrastive learning,
i.e., a Siamese neural network. 
The network takes a pair of callsite and callee embeddings as input, 
and generates their feature vectors,
which will be concatenated together and fed into a classifier
to calculate the difference score of the input pair. 

\vspace{-0.1cm}
\subsection{Workflow of the Learner}
\vspace{-0.1cm}
The input to the \textbf{Learner} is plenty of binaries, 
and outputs are models that could be used to embed program slices and report difference scores.
In total, there are 5 steps.

\begin{itemize}[leftmargin=.35cm,noitemsep,topsep=2pt]

    \item [\textbf{*}] \textbf{1: Collecting ground-truth callsite-callee pairs.}
    For dcalls, we simply extract callsite-callee pairs based on call instructions.
    For icalls, we dynamically run several testing programs with provided test suites and collect callsite-callee pairs at runtime.
    Specifically, we utilize Intel PT~\cite{PT} to collect traces for user-mode binaries
    and PANDA~\cite{dolan2013panda} for the Linux kernel.

    \item [\textbf{*}] \textbf{2: Statically extracting callsite-callee pair slices and functions from binaries.}
    With collected ground truths, we apply an inter-procedural slicing algorithm on binaries to extract slices for each callsite and its associated callee. 
    Meanwhile, we build a function dataset from training binaries to train an embedding model later.

    \item [\textbf{*}] \textbf{3: Slice preprocessing and embedding.}
    In this step, we symbolize instructions in the slices to reduce dimensions of data used in the following embedding model and neural network to make those models converge faster.
    Meanwhile, we train a doc2vec model using the collected function dataset. 
    The doc2vec model is then used to embed slices into vectors required by the neural network.

    \item [\textbf{*}] \textbf{4: Establishing a vectorized callsite-callee dataset.}
    In this step, we vectorize positive (matching) and negative (non-matching) callsite-callee pairs with the trained doc2vec model.
    Subsequently, we label positive ones as 1 and negative ones as 0.

    \item [\textbf{*}] \textbf{5: Training a Siamese neural network.}
    In this step, we construct a Siamese neural network with two parallel feature extraction layers,
    and train the network with the labeled dataset to produce difference scores.

\end{itemize}

\vspace{-0.1cm}
\subsection{Workflow of the Transfer Learning}
\vspace{-0.1cm}
With the proposed \textbf{Learner} framework, we perform transfer-learning between
\textbf{dLearner} and \textbf{iLearner}.

\begin{itemize}[leftmargin=.35cm,noitemsep,topsep=2pt]
    \item [\textbf{*}] \textbf{1: Pre-training the \textbf{dLearner}.}
    With collected binaries, we first train the \textbf{dLearner} with statically-extracted
    dcall pairs, following the standard train-validation-test procedure. After pre-training,
    we select the best-performance models for transfer learning.
    
    \item [\textbf{*}] \textbf{2: Initializing the \textbf{iLearner}.}
    We initialize parameters of models in \textbf{iLearner} with 
    values of corresponding parameters of the selected models, including the parameters
    of the doc2vec model and the Siamese network.
    
    \item [\textbf{*}] \textbf{3: Fine-tuning the \textbf{iLearner}.}
    Finally, we train the models of \textbf{iLearner} with dynamically-collected
    icall pairs after initialization, i.e., fine-tuning.

\end{itemize}

\vspace{-0.1cm}
\section{Methodology} \label{Design}
\vspace{-0.1cm}
We first introduce the contrastive \textbf{Learner} in detail,
i.e., context extraction, context embedding and callsite-callee matching,
and then describe the transfer learning.

\vspace{-0.1cm}
\subsection{Context Extraction via Slicing}
\vspace{-0.1cm}
Recent studies have shown that DNNs trained in a completely data-driven way without domain knowledge may be non-explainable and unpredictable, 
whose results may even conflict with prior expert knowledge. 
However, a system based completely on expert knowledge may have limitations in the scope and capability of solving problems, due to insufficient knowledge or improper inference logic. 

Therefore, we integrate expert knowledge into the deep learning system.
Specifically, we perform program slicing in advance.
The slicing step aims at using expert knowledge to preliminary extract useful information for matching callsite and callee pairs. 
Besides, shorter code gadgets after slicing are more favorable for embedding.

\begin{table}[t]
\setlength{\abovecaptionskip}{0cm}
\setlength{\belowcaptionskip}{0cm}
\footnotesize
\caption{Data passing rules in the calling convention of the System V AMD64 Application Binary Interface (ABI).}
\label{tab:data-passing}
\resizebox{0.48\textwidth}{!}{%
\begin{tabular}{@{}ccc@{}}
\toprule[1pt]
\textbf{Data Type}   & \textbf{Example}   & \textbf{Passing}  \\ \hline
\begin{tabular}[c]{@{}c@{}}INTEGER,\\ POINTER\end{tabular}        & \begin{tabular}[c]{@{}c@{}}\texttt{char}, \texttt{short},\\ \texttt{int}, \texttt{long}\end{tabular}   & \begin{tabular}[c]{@{}c@{}}Argument: \texttt{rdi}, \texttt{rsi}, \texttt{rdx}, \texttt{rcx}, \texttt{r8}, \texttt{r9}\\ Return value: \texttt{rax}, \texttt{rdx}\end{tabular} \\ \midrule
\begin{tabular}[c]{@{}c@{}}SSE,\\ SSEUP\end{tabular}              & \begin{tabular}[c]{@{}c@{}}\texttt{float},\\ \texttt{double}\end{tabular}            & \begin{tabular}[c]{@{}c@{}}Argument: \texttt{xmm0} to \texttt{xmm7}\\ Return value: \texttt{xmm0}, \texttt{xmm1}\end{tabular}             \\ \midrule
\begin{tabular}[c]{@{}c@{}}X87, X87UP,\\ COMPLEX\_X87\end{tabular} & \texttt{long double}                                                        & \begin{tabular}[c]{@{}c@{}}Argument: stack\\ Return value: \texttt{st0}, \texttt{st1}\end{tabular}                      \\ \midrule
MEMORY                                                            & \begin{tabular}[c]{@{}c@{}}struct, \\ array, union\end{tabular} & \begin{tabular}[c]{@{}c@{}}Argument: stack\\ Return value: (address in) \texttt{rax}\end{tabular}                \\ \bottomrule[1pt]
\end{tabular}%
}
\vspace{-2mm}
\end{table}

The principle of slicing is to identify and preserve instructions related to data dependencies between icallsites and icallees,
including local variables that passed between functions (arguments and return values) and global variables.
To get as much information as possible, we perform a depth-first traversal of all basic blocks in callsite and callee function's control-flow graph (CFG).
For global data dependencies, we keep instructions whose operands are related to values in the data segment.
For inter-procedural local data dependencies,
we keep those concerning stack memory and registers used for function arguments and return values,
based on rules of data passing~\cite{lu2018system} shown in Table~\ref{tab:data-passing}.
To be conservative, we do not drop control-flow instructions.
Details of slicing algorithms are presented in Section~\ref{algo:slicing}.

\vspace{-0.1cm}
\subsection{Context Embedding}
\vspace{-0.1cm}

Required by most neural networks, inputs need to be embedded into vectors or tensors. 
Therefore, we adopt doc2vec, a common approach in the field of NLP, to embed slices.

Before embedding, instructions should be tokenized to avoid nonexistent tokens caused by punctuation. 
For instance, instruction 
\texttt{mov rax, [rdi]}
should be tokenized into 
\texttt{"mov", "rax", ",", "[", "rdi", "]"}.
Moreover, instructions from a fresh binary
may have tokens unseen in the trained doc2vec model,
known as the Out-of-Vocabulary (OOV) phenomenon.
Thus we need to symbolize slices before embedding. 

\subsubsection{Symbolization}
The general idea of symbolization is to replace open-set tokens with closed-set tokens. 
Open-set tokens are tokens that can have many variants, 
including immediate operands, user-defined function names, user-defined variables, and so on.
Contrastively, closed-set tokens refer to tokens that have limited variants.
For example, \texttt{20h} is an open-set token in instruction \texttt{mov eax, 20h}. 
It can be replaced by \texttt{num}, which is a closed-set token.

\begin{table*}[ht]
\setlength{\abovecaptionskip}{0cm}
\setlength{\belowcaptionskip}{0cm}
    \centering
    \caption{Symbolization Rules.}
    \label{tab:sym}
    \resizebox{\textwidth}{!}{
    \begin{tabular}{c|cccccccccccc}
    \toprule[1pt]
    Symbolization & loc\_ABCD   & arg\_ABCD   & sub\_ABCD   & var\_ABCD   & struct\_ABCD   & unk\_ABCD   & byte\_ABCD   & off\_ABCD      & *word\_ABCD   & flt\_ABCD   & dbl\_ABCD   & a\_String        \\ \hline
    Strict        & loc        & arg        & fun        & var        & struct        & unk        & byte        & offset        & word        & flt        & dbl        & str             \\
    Loose         & loc+ABCD$\%$N & arg+ABCD$\%$N & fun+ABCD$\%$N & var+ABCD$\%$N & struct+ABCD$\%$N & unk+ABCD$\%$N & byte+ABCD$\%$N & offset+ABCD$\%$N & *word+ABCD$\%$N & flt+ABCD$\%$N & dbl+ABCD$\%$N & str+len(String) \\ %
    \bottomrule[1pt]
    \end{tabular} 
}
\vspace{-3mm}
\end{table*}

Further, the intensity of symbolization should be taken into account. 
We compare two symbolization policies: {\em strict} symbolization and {\em loose} symbolization. 
By strict, it means that the symbolization process transforms open-set tokens in the same kind into a single closed-set token.
For instance, given an open set of user-defined function names \texttt{$foo\_{0}$, $foo\_{1}$,...,$foo\_{\infty}$}, 
any token in it will be replaced by the same closed-set token \texttt{fun}. 
Strict symbolization is the most commonly used policy in preprocessing, because it can eliminate OOV. 
However, strict symbolization may lose data-flow information, 
which often contributes to the determination of the function call targets.
For example, strict symbolization turn all strings into one token \texttt{"str"}.

Hence we propose {\em loose symbolization} to preserve data-flow information and meanwhile maintain a finite-size token corpus. 
Through modulo arithmetic, an open set like \{$foo\_{0}$, $foo\_{1}$,...,$foo\_{\infty}$\} can be transformed into \{$foo\_{0}$, $foo\_{1}$,...,$foo\_{(N-1)}$\} where N is a hyperparameter. 
As for strings, we simply take the length of a string as a suffix, and replace the string with \texttt{str\_len}. 
Additionally, several kinds of tokens are symbolized according to their semantics. 
For example, operands of a dcall instruction are considered to be a function, and thus we replace them with \texttt{"fun"}.
Detailed rules of symbolization are summarized in Table~\ref{tab:sym}.

\subsubsection{Vectorization} 
After symbolization, \sysname adopts doc2vec, a popular model used in NLP, to embed slices into vectors. 
A doc2vec model takes paragraphs of tokens as input and calculates the distributions of both paragraphs and tokens. 
To capture the semantic information of low-frequency tokens, 
we choose the Distributed Bag of Words of Paragraph Vector (PV-DBOW) model~\cite{le2014distributed}, 
and adjust it to apply to assembly language.
Note that, compared with word2vec and PalmTree~\cite{li2021palmtree}
(\added{For detailed evaluation of different embedding techniques, 
please refer to Appendix~\ref{appendix:embedding}.}), 
doc2vec is able to calculate the word embedding and paragraph embedding at the same time, and the paragraph embedding is shared during multiple training of word embeddings in one paragraph. Thus the generated word embedding in fact involved both inter-token and inter-instruction information.

Formally, an $m$-token callsite slice $\vec{\pi_i} = \{u_0, u_1,...,u_m | u \in R^j\}$ 
and an $n$-token callee slice $\vec{\alpha_i} = \{t_0, t_1,...,t_n | t \in R^j\}$
are mapped into 
\[
\textit{G}(\vec{\pi_i}) \rightarrow \vec{Q_i} = \{\vec{E}_{u_0}, \vec{E}_{u_1},...,\vec{E}_{u_m} | \vec{E} \in R^k\}, and
\]
\[
\textit{G}(\vec{\alpha_i}) \rightarrow \vec{A_i} = \{\vec{E}_{t_0}, \vec{E}_{t_1},...,\vec{E}_{t_n} | \vec{E} \in R^k\}
\]
where \textit{G} is the doc2vec model as a mapping 
$\textit{G}: \textbf{X} \rightarrow \textbf{Z}$ 
between the token space $\textbf{X}:R^j$ and the embedding space $\textbf{Z}:R^k$.
Note that embeddings for each token in a paragraph are concatenated together, i.e.
$\vec{Q_i} \leftarrow \vec{E}_{u_0}\oplus\vec{E}_{u_1}\oplus...\vec{E}_{u_m}$;
$\vec{A_i} \leftarrow \vec{E}_{t_0}\oplus\vec{E}_{t_1}\oplus...\vec{E}_{t_n}$.

However, doc2vec is designed to be applied to natural languages (e.g., English). 
But the prior knowledge of natural languages is quite different from the assembly. 
Thus \sysname adjusts two parameters of doc2vec intuitively. 

\begin{itemize}[leftmargin=.35cm,noitemsep,topsep=0pt]
    \item \texttt{sample}: 
    In natural languages, high-frequency tokens are mostly function words.
    Therefore, these tokens are usually downsampled to reduce their frequency. 
    Yet high-frequency tokens in assembly language can carry much information 
    (e.g., \texttt{comma} to distinguish operands). 
    As a result, we do not downsample high-frequency tokens.
    \item   \texttt{min\_count}: 
    Low-frequency words caused by wrong segmentation results of sentences 
    are often ignored during training an embedding model of natural languages.
    On the contrary, low-frequency tokens in program analysis scenarios can be semantically deterministic.
    Hence we set the \texttt{min\_count} parameter to 0. 
\end{itemize}

\vspace{-0.1cm}
\subsection{Structure of the Matching Network}
\vspace{-0.1cm}
For embedded callsites and callees,
we further build a Siamese neural network to predicate their difference scores.

An embedded callsite slice $\vec{Q_i}$ will pass through feature extraction layers $\phi$ that output a feature vector $\vec{q_i} = \{\phi(\vec{Q_i}) | \phi: \textbf{Z} \rightarrow \textbf{F}\}$,
where $\textbf{F}:R^f$ is the feature space.
Similarly, for an embedded callee slice $\vec{A_i}$ we can obtain a feature vector 
$\vec{a_i} = \{\phi'(\vec{A_i}) | \phi': \textbf{Z} \rightarrow \textbf{F}\}$
with another set of feature extraction layers $\phi'$.
Then to calculate the matching score, we concatenate two feature vectors together, 
considering that currently there is no theoretical proof of which distance measure is optimal for feature vectors. In other words, different data/scenarios may need different distance measures. 
Therefore we utilize a fully-connected network (FCN) to predict a score with the concatenated vector, 
i.e., "let the data talk". The FCN $\sigma$ is essentially an adaptive (trainable) "distance":
$d_i = \sigma(\vec{q_i}\oplus\vec{a_i})$.

The contrastive loss~\cite{hadsell2006dimensionality} is used as the optimization goal of our Siamese network:
\vspace{-0.1cm}
\[
L=\frac{1}{2N}\sum_{i=1}^{N}[y_{i}d^{2}_{i}+(1-y_{i})\max \{1-d_{i},0\}^{2}]
\]
where $N$ is the number of input pairs, $y_{i}$ (i.e., 1 or 0) is the label of the input pair (i.e., match or not). 
The optimization goal indicates that, 
if the input pair match ($y_{i}=1$), then the output (difference score) $d_i$ should be close to 0;
otherwise, the output should be close to 1.

According to the output $d$, we can set a threshold to determine whether the callsite and callee match:
\[
matching=
\begin{cases}
yes     & \text{$d$ \textless threshold}\\
no      & \text{otherwise}
\end{cases}
\]

\vspace{-0.1cm}
\subsection{Transfer Learning}
\vspace{-0.1cm}
With the proposed \textbf{Learner} framework,
we utilize a two-stage transfer-learning training mechanism,
i.e., pre-training with dcalls and fine-tuning with icalls.
Specifically, given two Siamese neural networks
\vspace{-0.1cm}
\[
\Lambda_d = \sigma_d(\phi_d(\vec{Q_i}, \theta_d)\oplus\phi'_d(\vec{A_i}, \theta'_d))
\]
\vspace{-0.1cm}
and
\vspace{-0.1cm}
\[
\Lambda_i = \sigma_i(\phi_i(\vec{Q_i}, \theta_i)\oplus\phi'_i(\vec{A_i}, \theta'_i))
\]
\vspace{-0.1cm}
for dcalls and icalls respectively,
where $\theta$ indicates parameters of $\phi$. 
we first train $\Lambda_d$ with dcall pairs, 
and then initialize $\phi_i$ and $\phi'_i$ with $\phi_d$ and $\phi'_d$,
and further fine-tune $\Lambda_i$ with icall pairs.
Note that $\sigma_i$ is trained from scratch,
and the doc2vec model follows the same training mechanism.

\vspace{0.2cm}
\section{Implementation} \label{Implementation}
\vspace{-0.1cm}

\vspace{-0.1cm}
\subsection{Dataset Collection}
\vspace{-0.1cm}
The datasets that \sysname used require two kinds of data: 
assembly functions for training the doc2vec model and callsite-callee pairs for training the Siamese neural network.

\vspace{-0.1cm}
\subsubsection{Functions}
In analogy with natural languages, we regard functions as the "paragraphs", instructions as "sentences", opcodes and operands as "words",
and train a doc2vec model to embed slices into vectors.

We write a Python script for IDA Pro to extract functions from binaries.
Note that only functions in the \texttt{.text} section are extracted. 
As for those in other sections, we have to identify which shared libraries they are in. 
All involved shared libraries are analyzed later to extract their functions.

\vspace{-0.1cm}
\subsubsection{Callsite-callee pairs}
The primary goal is to record addresses of callsite-callee pairs in binaries. 

Direct-call pairs can be easily obtained with IDA Pro by simply traversing binaries and recording addresses of callsites and callees. 
For icalls, 
we utilize dynamic anaylses to collect ground truths.
For user-mode binaries, we instrument all icallsites with an LLVM pass to output the callee at runtime. With coverage-guided fuzzers such as AFL \cite{AFL} and program test suites as fuzzing seeds, we can cover most functional code.
After fuzzing, the indirect callsite-callee pairs are collected by running the program with generated inputs.
For the kernel, we emulate it in PANDA~\cite{dolan2015repeatable}.
By parsing emulation logs, we can obtain the icall pairs. 
For more details, please refer to Appendix \ref{appendix:datacollection}.

\vspace{-0.1cm}
\subsection{Slicing} \label{algo:slicing}
\vspace{-0.1cm}
We implement the slicing algorithm with the IDAPython \cite{idapy} SDK provided by IDA Pro. 
Before slicing, we filter out cases where IDA Pro fails or goes wrong.

\begin{algorithm}[t]
\scriptsize
\DontPrintSemicolon
\caption{Slicing of a callsite}
\label{alg:concallsite}
\KwIn{CallsiteSet}
\KwOut{CallsiteResult}

$\rm CallsiteResult \;{\leftarrow}\; \{\} $\;
\ForEach{Callsite {\bf in} CallsiteSet}{
    $\rm StackSet, RegSet, GlobalVarSet, CrtlFlowSet \;{\leftarrow}\; \{\} $\;
    \For{$Insn\;{\leftarrow}\;FuncStart : Callsite$}{
        \eIf{isStackInsn(Insn)}{
            $\rm StackSet \;{\leftarrow}\; StackSet \cup \{Insn\}$\;
        }{
            \ForEach{Op {\bf in} InsnOperands}{
                \If{isArgRegInOp(Op)}{
                    $\rm RegSet \;{\leftarrow}\;RegSet \cup \{Insn\}$\;
                }
            }
        } 
    }
    \For{$Insn\;{\leftarrow}\;Callsite :FuncEnd$}{
        \ForEach{Op {\bf in} InsnOperands}{
            \If{isRetRegInOp(Op)}{
                $\rm RegSet \;{\leftarrow}\;  RegSet \cup \{Insn\}$\;
            }
        }
    }
    $\rm GlobalVarSet \;{\leftarrow}\;  getGlobalVarXref(Function)$\;
    $\rm CrtlFlowSet \;{\leftarrow}\;  getCrtlFlowInsn(Function)$\;
    $\rm SliceResult \;{\leftarrow}\; StackSet \cup  RegSet\cup GlobalVarSet \cup CrtlFlowSet $\;
    $\rm CallsiteResult \;{\leftarrow}\; CallsiteResult \cup \{SliceResult\}$\;
}

\KwRet{CallsiteResult}
\end{algorithm}

\begin{algorithm}[t]
\scriptsize
\DontPrintSemicolon
\caption{Slicing of a callee}
\label{alg:concallee}
\KwIn{CalleeSet}
\KwOut{CalleeResult}

$\rm CalleeResult \;{\leftarrow}\; \{\} $\;
\ForEach{Callee {\bf in} CalleeSet}{
    $\rm Function\;{\leftarrow}\;makeFunction(Callee)$\;
    $\rm StackSet, RegSet, GlobalVarSet, CrtlFlowSet \;{\leftarrow}\; \{\} $\;
    \For{$Insn\;{\leftarrow}\;FuncStart : FuncEnd$}{
        \eIf{isStackInsn(Insn)}{
            $\rm StackSet \;{\leftarrow}\; StackSet \cup \{Insn\}$\;
        }{
            \ForEach{Op {\bf in} InsnOperands}{
                \uIf{isArgRegInOp(Op)}{
                    $\rm RegSet \;{\leftarrow}\; RegSet \cup \{Insn\}$\
                }\ElseIf{isRetRegInOp(Op)}{
                    $\rm RegSet \;{\leftarrow}\; RegSet \cup \{Insn\}$\;
                }
            }
        } 
    }
    $\rm GlobalVarSet \;{\leftarrow}\;  getGlobalVarXref(Function)$\;
    $\rm CrtlFlowSet \;{\leftarrow}\;   getCrtlFlowInsn(Function)$\;
    $\rm SliceResult \;{\leftarrow}\; StackSet \cup  RegSet \cup GlobalVarSet \cup CrtlFlowSet $\;
    $\rm CalleeResult \;{\leftarrow}\; CalleeResult \cup \{SliceResult\}$\;
}
\KwRet{CalleeResult}
\end{algorithm}

We extract slices from callsites (Algorithm \ref{alg:concallsite}) and callees (Algorithm \ref{alg:concallee}), then combine them according to the requirements of training or testing. 
First, we get the function where the callsite or callee address is located.
Since the function boundary of a target function called in an indirect way may not be correctly recognized by static analysis,
we force callee addresses to be starts of functions when slicing callees. 
Then, we walk through instructions of the function, deciding whether to keep them based on operands. 
To preserve local variables' inter-procedural data dependencies, we identify and retain the information about function signatures. 
For arguments, we extract instructions concerning stack memory and registers used for arguments from the first half of the callsite function (i.e., instructions before this \texttt{call} instruction) and the whole callee function. 
For return values, we extract instructions containing registers used for return values from the second half of the callsite function (i.e., instructions after this \texttt{call} instruction) and in callees. 
To preserve global variables' data dependencies, we get cross-reference instructions of global variables in both callsite and callee functions. 
Finally, we gather control-flow instructions, 
and the union of those parts is taken as the result.

\vspace{-0.1cm}
\subsection{Embedding}
\vspace{-0.1cm}
\sysname utilizes IDA Pro to disassemble instructions, 
so we take advantage of its naming rules to symbolize instructions. 
By default, data structures are named according to their addresses. 
For example, a user-defined function at address \texttt{0x43B9D0} in the \texttt{.text} section is named as \texttt{sub\_43B9D0}. 
Therefore, we can symbolize the function as \texttt{fun} (strict) or \texttt{func0} (loose), 
assuming that the hyper-parameter N is set to 10. 
As shown in Table~\ref{tab:sym}, we consider 12 situations in total.

\vspace{-0.1cm}
\section{Evaluation} \label{Evaluation}
\vspace{-0.1cm}
We evaluate \sysname from the following aspects:
\begin{itemize}
    \item {\bf Performance of icallee recognition.} 
    We compare \sysname with SOTA solutions, conduct ablation studies,
    and discuss its generalization and time efficiency.
    \item {\bf Applications of \sysname.}
    We apply \sysname to binary similarity detection and hybrid fuzzing to examine
    whether it can promote their performance.
    \item {\bf Interpretability of \sysname.}
    We interpret the \textbf{Learner} framework used by \sysname.

\end{itemize}

\vspace{-0.1cm}
\subsection{Evaluation Setup}
\vspace{-0.1cm}
Experiments are performed on a machine equipped with Ubuntu 18.04 LTS. 
The machine has an Intel CPU (Intel(R) Xeon(R) Gold 6248R CPU @ 3.00GHz), four NVIDIA GPUs (A100 PCIE) and 768GB RAM, 
and is installed with LLVM 12.0.1, GCC 7.5.0, libipt 2.0.0 (commit \texttt{892e12c5}), a docker image of PANDA (git tag: \texttt{0729fd0d}), IDA Pro 7.6, and Python 3.6.9.
The Python is equipped with gensim 4.2.0 and PyTorch 1.9.0. 

\begin{table}[h]
\centering
\setlength{\abovecaptionskip}{0cm}
\setlength{\belowcaptionskip}{0cm}
\caption{Dataset Statistics.}
\label{tab:datasets}
\begin{tabular}{c|cccc}
\toprule[1pt]
Dataset & \# Projects & \# Binaries & \# Functions & \# Pairs \\ \hline
Direct Call   & 19K    & 261K  & 68M & 406M   \\
Indirect Call   & 52       & 183     & 343K & 30K \\ 
GNU Binutils$^*$    & 1       & 694   & 963K    & 5M \\ \bottomrule[1pt]
\multicolumn{5}{l}{\footnotesize * For cross-compiler and cross-version evaluation.}
\end{tabular}
\vspace{-0.3cm}
\end{table}

\subsubsection{Datasets} \label{sssec:datasets}
Table \ref{tab:datasets} shows statistics on the number of projects, binaries, functions 
and callsite-callee pairs of the datasets we collect.
All binaries are in the x64 architecture.
For the dcall dataset, we first build enormous binaries automatically with the apt package manager
and then extract functions and direct callsite-callee pairs.
For the icall dataset, we collect binaries rich in icalls,
including the Linux kernel (v5.3.11), the Firefox browser (v72.0a1),
and corresponding shared libraries.
After dynamic testing, we extract functions from them and perform the slicing. 
To build a balanced dataset, we set the ratio of positive pairs to negative pairs to 1:1
and assemble negative pairs by randomly choosing unmatched callsites and callees from the ground truths.
To avoid negative pairs that are actually positive pairs not covered by dynamic testing, we additionally check the source-level type of the unmatched pairs with the help of debug information, which contains type information of function calls.
Note that different projects usually have different contributors, whose coding styles can be varied,
e.g., Firefox has over 100 contributors in the last 90 days~\cite{firefoxauthors},
and thus we believe the datasets are diversified based on the number of projects.

\begin{figure}[!]
    \centering
    \subfigure[Callee Distribution]{
    \includegraphics[width=3.5cm]{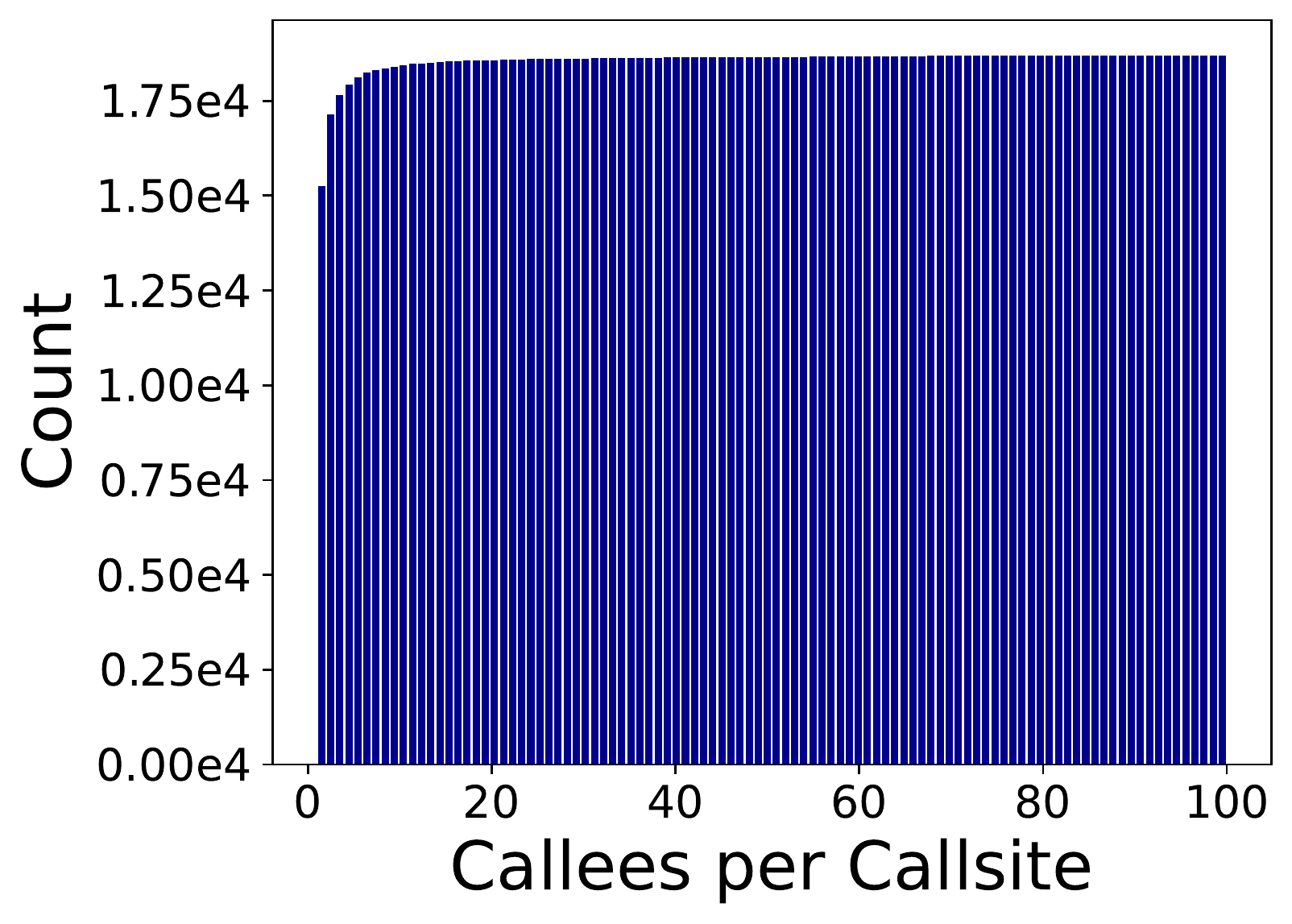}
    }
    \quad
    \subfigure[Callsite Distribution]{
    \includegraphics[width=3.5cm]{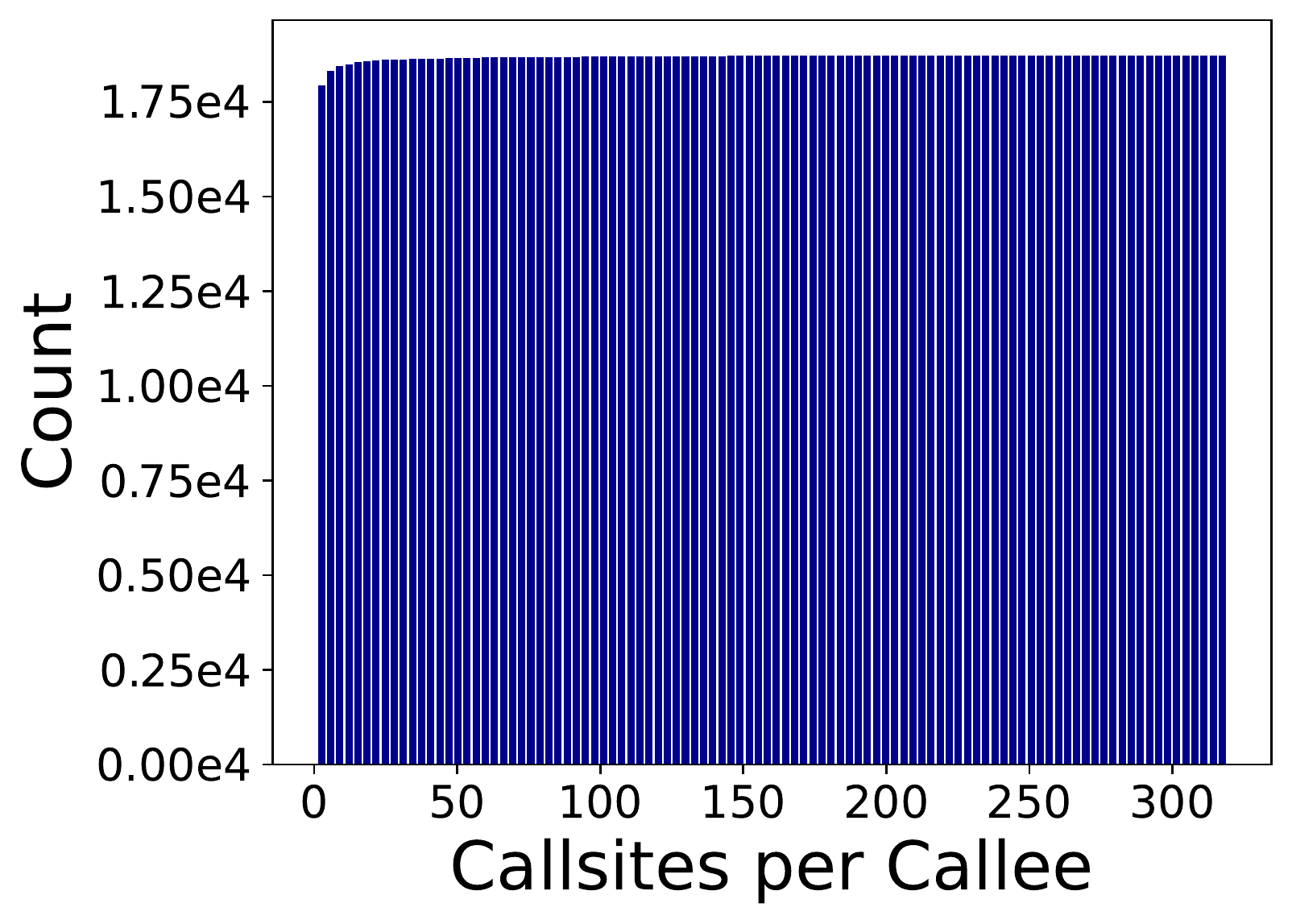}
    }
    \caption{Distribution of callees per callsite (left) and callsites per callee (right) in the icall dataset.}
    \label{fig:distribution}
    \vspace{-0.3cm}
\end{figure}

Additionally, we study the distribution of callsites and callees in the icall dataset.
\deleted{with the Cumulative Distribution Function (CDF) plot.}
For a malformed dataset whose callsites generally have the same small set of callees,
almost any algorithm will do well by just guessing those callees the majority of the time.
As shown in Figure~\ref{fig:distribution}, 
the number of "callsites per callee" is small for the majority of callees, 
indicating that \added{the callsites with common callees will be few.}
\deleted{it is unlikely for most callsites to invoke the same small set of callees.}
And the number of "callees per callsite" is small for the majority of callsites, further demonstrating the diversity of the dataset.

\begin{table*}[t]
\centering
\setlength{\abovecaptionskip}{0cm}
\setlength{\belowcaptionskip}{0cm}
\caption{Performance of \sysname (in bold) on the icall dataset. Results not in bold are presented for ablation studies.}
\label{tab:ablation}
\begin{tabular}{c|c|c|c|c|ccc|ccc}
\toprule[1pt]
\multirow{2}{*}{Setting} & \multirow{2}{*}{Context} & \multirow{2}{*}{Symbolization} & \multirow{2}{*}{\tabincell{c}{Siamese\\Network} } & \multirow{2}{*}{Mode}      & \multicolumn{3}{c}{Train}                     & \multicolumn{3}{c}{Test}                      \\ \cline{6-11} 
                         &                          &                                &                                  &                            & Precision     & Recall        & F1            & Precision     & Recall        & F1            \\ \hline
0                        & Sliced                   & Loose                          & FCN                              & \textit{dcall}             & 93.4\%          & 87.9\%          & 90.6\%          & 93.8\%          & 87.5\%          & 90.5\%          \\
1                        & Sliced                   & Loose                          & FCN                              & \textit{icall}             & 76.8\%          & 75.6\%          & 76.2\%          & 70.3\%          & 63.7\%          & 66.8\%          \\
\textbf{2}               & \textbf{Sliced}          & \textbf{Loose}                 & \textbf{FCN}                     & \textit{\textbf{transfer}} & \textbf{99.2\%} & \textbf{96.8\%} & \textbf{98.0\%} & \textbf{97.3\%} & \textbf{90.9\%} & \textbf{94.6\%} \\
3                        & Sliced                   & Loose                          & FCN                              & \textit{zero-shot}         & -               & -               & -               & 93.0\%          & 85.9\%          & 89.3\%          \\
4                        & \textit{Full}            & Loose                          & FCN                              & icall                      & 74.1\%          & 72.9\%          & 73.5\%          & 57.4\%          & 53.0\%          & 55.1\%          \\
5                        & Sliced                   & \textit{Strict}                & FCN                              & icall                      & 75.3\%          & 74.2\%          & 74.7\%          & 61.9\%          & 56.6\%          & 59.1\%          \\
6                        & Sliced                   & Loose                          & \textit{LSTM}                    & icall                      & 71.0\%          & 70.1\%          & 70.5\%          & 67.8\%          & 61.5\%          & 64.5\%          \\
7                        & Sliced                   & Loose                          & \textit{TextCNN}                 & icall                      & 73.7\%          & 72.3\%          & 73.0\%          & 69.5\%          & 63.0\%          & 66.1\%          \\
8                        & Sliced                   & Loose                          & \textit{1dCNN}                   & icall                      & 77.3\%          & 75.7\%          & 76.5\%          & 68.4\%          & 61.7\%          & 64.9\%          \\ 
\bottomrule[1pt]
\end{tabular}
\end{table*}

\textbf{Dataset split.}
A common split method is cross-validation: randomly choosing, e.g., 70\% pairs for training, 20\% for validation and 10\% for testing, without considering the distribution of the data (e.g., the originating binaries).
But it can lead to severe overfitting issues, i.e., the model overfits patterns of data from binaries 
in the dataset and cannot generalize to data from binaries outside the dataset.

We have conducted an experiment following this split method.
The final F1 scores of the model on the icall dataset are 98.9\% for training and 94.6\% for testing.
However, when we apply the trained model to data extracted from binaries outside the dataset, 
the F1 drops sharply to about 53.7\%, indicating that the trained model's 
generalization ability is poor. In other words, the model overfits the dataset.

To acquire a better generalization ability across binaries,
{\em we extract pairs from different binaries for training and testing to evaluate the generalization performance across different binaries.}
Therefore, we first randomly choose 80$\%$ of the binaries for training, 10$\%$ for validation, and $10\%$ for testing. Then pairs are further extracted from these binaries.
Since the dataset consists of binaries by different authors,
thus there is little shared code across the split datasets.

\vspace{-0.2cm}
\subsubsection{Hyperparameters} 
We set the \texttt{batch\_size} to 512, and train the network 20 epochs. 
The optimizer is \texttt{rmsprop}, the learning rate is 0.001, 
the threshold for the final decision is 0.5,
and the embedding dimension of the doc2vec model is 100. 
The final classifier network of the Siamese neural network is an FCN 
consisting of three layers with 512, 512, and 1 neuron(s) respectively.
The sigmoid function is used as the final activation function.
We adopt Batch Normalization \cite{ioffe2015batch} and Dropout \cite{srivastava2014dropout} to help the network converge, and the dropout rate is set to 0.2.
The hyper-parameter N of loose symbolization is set to 10. 
Note that, these hyperparameters are selected based on several rounds of dry-run experiments.

\vspace{-0.2cm}
\subsubsection{Evaluation Metrics} \label{sssec:metrics}
We choose the common metrics \textit{Precision}, \textit{Recall} and \textit{F1-Measure} (\textit{F1}) to evaluate the performance of models. 
These metrics are computed from the number of \textit{True Positives} (\textit{TP}), 
\textit{True Negatives} (\textit{TN}),
\textit{False Positives} (\textit{FP}), and
\textit{False Negatives} (\textit{FN}).
An FP is a pair classified as match but actually does not match.
An FN is a pair classified as unmatch but actually matches.

\vspace{-0.1cm}
\subsection{Performance of \sysname}
\vspace{-0.1cm}
\subsubsection{Overall Performance}
Overall, we choose the loose symbolization method and FCN feature extraction layers
and train the Siamese neural network on sliced contexts with the transfer-learning technique. 
As shown in Table~\ref{tab:ablation}, \sysname has an \textit{F1} of 94.6\%, recall of 90.9\%, and precision of 97.3\%.

\vspace{-0.1cm}
\subsubsection{Comparison with state-of-the-art solutions} \label{sssec:sotaeval}
We compared \sysname with several closely relevant solutions which recognize icallees as well. 
Since the refinement of TypeArmor
fails to discuss their precision/recall in recognizing icallees and has not open-sourced yet,
we only compare \sysname with BPA and TypeArmor
as well as popular binary analysis tools such as IDA Pro, Angr and GHIDRA.
We use the same binaries as BPA: 
the SPEC CPU 2006 benchmark and 4 server applications
(memcached-1.5.4, lighttpd-1.4.48, exim-4.89, and nginx-1.10).

\begin{figure}[h]
    \centering
    \includegraphics[width =.4\textwidth]{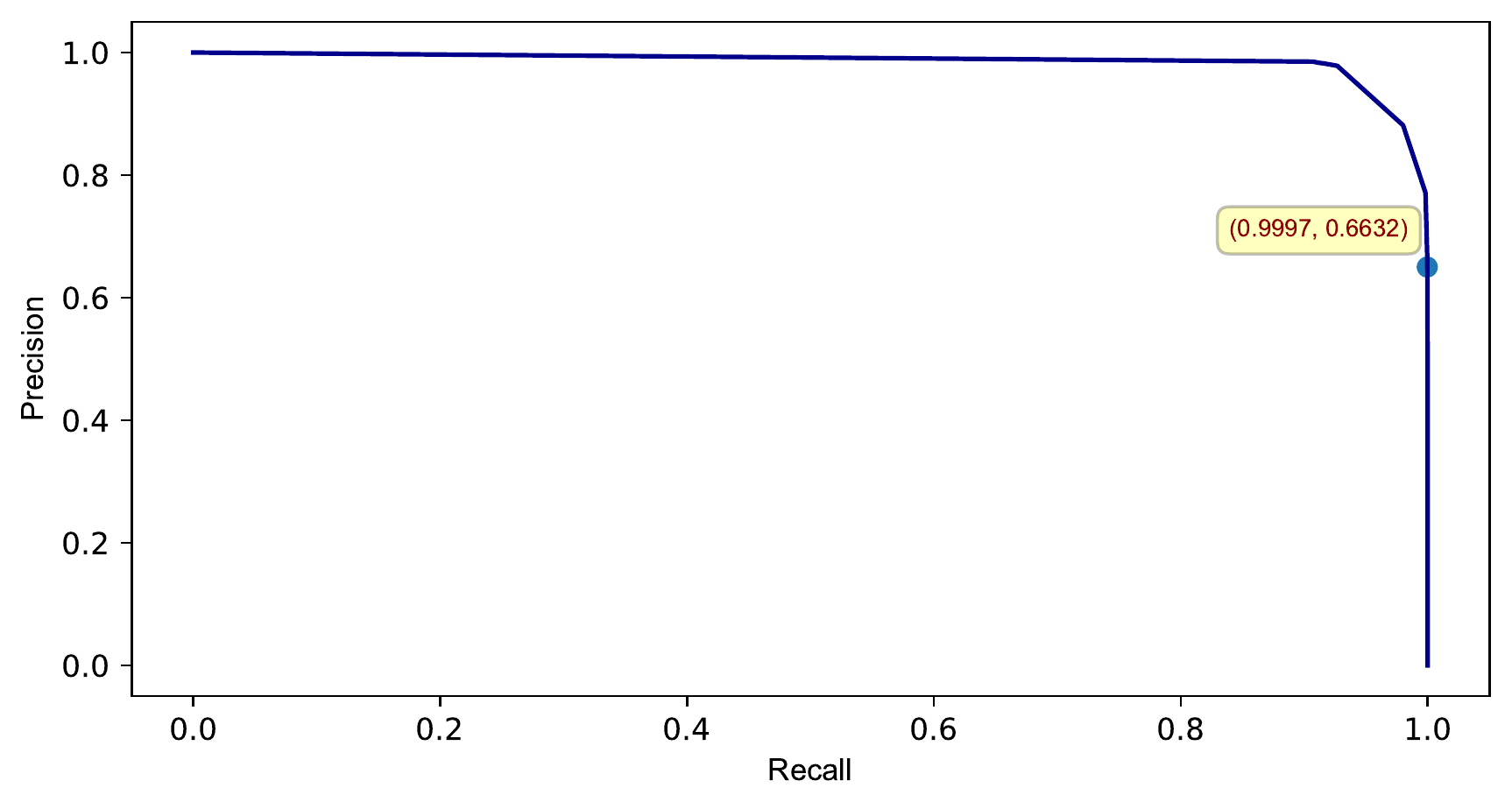}
    \setlength{\abovecaptionskip}{0cm}
    \caption{Precision-Recall Curve of \sysname.}
    \label{fig:pr}
     \vspace{-0.2cm}
\end{figure}

Since BPA is not open-sourced, we adopt the results from their paper:
based on a dynamically collected dataset~\cite{kim2021icall},
BPA and TypeArmor have precision of 57.6\% and 35.1\%,
recall of \added{100}\deleted{99.9}\% and \added{99.9}\deleted{100}\%, and thus F1-measures of 73.1\% and 51.9\% respectively.
For fair comparison, we report \sysname's precision-recall (PR) curve in Figure~\ref{fig:pr}.
As shown, the precision drops as the recall increases, and the precision remains 66\% when recall reaches 99.9\%.
As for real-world binary analysis tools such as IDA Pro, Angr, GHIDRA, etc.,
they identify icall targets by constant propagation. 
Although constant propagation can avoid false positives, i.e. has a 100\% precision rate, 
it can only resolve very few targets and has high false negatives, i.e. has a recall rate close to zero,
and thus has an F1-measure close to 50\%.
For icallsites of subject binaries in Table~\ref{tab:ICT}, 
constant propagation can at most recognize 8 targets in \texttt{403.gcc},
and cannot recognize any target(s) in over half of the binaries.

We also calculate the average indirect call target (AICT) metric that TypeArmor and BPA used,
and the results are shown in Table \ref{tab:ICT}.
Additionally, we include a source-level type analysis solution LLVM-CFI~\cite{tice2014enforcing} as a reference.
Column \textbf{\#Functions} indicates the number of all functions in a binary, and
columns \textbf{\#iCallsites} and \textbf{\#AT} indicate the number of icallsites and address-taken functions respectively.
\added{For fair comparison, we use the set of icallees when \sysname's recall is 99.9\%.}
We assume the recovering results of TypeArmor are absolutely correct, 
though the accuracy of TypeArmor in identifying argument numbers is about 83$\%$, 
and much lower in identifying the usage of return value (less than 20$\%$).
Nonetheless, it shows that \sysname has smaller AICTs than state-of-the-art solutions on most binaries, 
and can reduce 40.1$\%$ icallees than TypeArmor on average,
which is better than BPA and the refinement solution.
While as expected, LLVM-CFI still outperforms \sysname,
since it is a source-level solution which could utilize function type information.

\begin{table}[t]
\footnotesize
\setlength{\abovecaptionskip}{0cm}
\setlength{\belowcaptionskip}{0cm}
\caption{AICT evaluation results. \textbf{\#AT} indicates numbers of address-taken functions and \textbf{\#CP} indicates numbers of callees found by constant propagation.}
\label{tab:ICT}
\resizebox{.48\textwidth}{!}{
\begin{tabular}{c|c|c|c|c|cccc}
\toprule[1pt]
\multirow{2}{*}{Binary} & \multirow{2}{*}{\#Functions} & \multirow{2}{*}{\#iCallsites} & \multirow{2}{*}{\#AT} & \multirow{2}{*}{\#CP} & \multicolumn{4}{c}{AICT}                             \\ \cline{6-9} 
                        &                        &                            &                     &                     & TypeArmor    & BPA          & \sysname        & LLVM-CFI  \\ \hline
nginx                   & 1118                   & 220                        & 744                 & 4                   & 420.5     & 525.1   & 383.0   & 21.5  \\
lighttpd                & 360                    & 56                         & 279                 & 0                   & 24.7      & 33.9    & 31.7    & 7.0   \\
exim                    & 622                    & 78                         & 344                 & 0                   & 38.0      & 30.6    & 22.4    & 5.7   \\
memcached               & 244                    & 50                         & 109                 & 0                   & 21.6      & 1.4     & 11.3    & 1.1   \\
400.perlbench           & 1793                   & 117                        & 664                 & 6                   & 536.6     & 363.7   & 354.0   & 24.0  \\
401.bzip2               & 79                     & 22                         & 2                   & 0                   & 1.0       & 2.0     & 1.4     & 1.0   \\
403.gcc                 & 4678                   & 44                         & 1050                & 8                   & 581.3     & 427.8   & 338.0   & 9.3   \\
433.milc                & 245                    & 6                          & 3                   & 0                   & 2.0       & 2.0     & 2.0     & 2.0   \\
445.gobmk               & 2537                   & 46                         & 1672                & 1                   & 1,413.3   & 1,297.2 & 672.4   & 600.9 \\
456.hmmer               & 506                    & 12                         & 20                  & 1                   & 22.0      & 2.8     & 7.2     & 10.0  \\
458.sjeng               & 145                    & 3                          & 8                   & 0                   & 7.0       & 7.0     & 7.0     & 7.0   \\
464.h264ref             & 533                    & 354                        & 40                  & 0                   & 28.9      & 26.4    & 20.9    & 2.1   \\
482.sphinx              & 336                    & 10                         & 7                   & 0                   & 1.9       & 0.7     & 5.6     & 5.0   \\ \hline
Average                 & 1,015.1                & 78.3                       & 380.2               & 1.5                 & \added{-}\deleted{238.4}     & \added{-}\deleted{209.3}   & \added{-}\deleted{142.8}   & \added{-}\deleted{53.6}  \\
\bottomrule[1pt]
\end{tabular}
}
\end{table}

\vspace{-0.1cm}
\subsubsection{Ablation Studies} \label{sssec:ablation}
To evaluate how key parts of \sysname influence the performance,
we perform ablation studies of transfer learning, slicing, symbolization and feature extraction layers of the Siamese network.

\textbf{Effect of Transfer learning.}
To evaluate the effect of transfer learning, we perform model training with 4 modes:
training and testing with dcall and icall datasets respectively, 
training on dcall dataset first and fine-tuning 
with icall dataset (i.e. transfer-learning), 
and training on dcall dataset and testing on icall dataset (i.e. zero-shot learning).
Table \ref{tab:ablation} shows that merely training with icall dataset can only achieve a 66.8\% \textit{F1} on the test set, and meanwhile suffers from over-fitting (\textit{F1} drops 9.4\% from training to testing).
While transfer-learning can boost the \textit{F1} during testing to over 94\%.
Even in the zero-shot learning setting, where we test the pre-trained dcall model with
the icall dataset without fine-tune, the \textit{F1} can still reach 89\%,
indicating that dcall and icall pairs can share many common patterns,
and thereby transfer-learning can greatly improve \sysname's performance.

\textbf{Effect of Slicing.}
To evaluate the effect of slicing, we first fixate other parts of \sysname. 
Based on the icall dataset, we compare two situations: full context and sliced context 
(Settings 1, 4 in Table~\ref{tab:ablation}).
As shown, the model trained with full context suffers from severe over-fitting: \textit{F1} drops 18.4\% from training to testing,
showing that processing binaries with slicing could greatly help the Siamese network comprehend the context.
It also indicates that
full contexts of one binary can significantly differ from those in another binary,
considering that different binaries in the icall dataset are compiled with different compilers 
and there is manually written assembly code in the Linux kernel.
Therefore the network overfits code patterns in training binaries.
Whereas performing slicing can "uniform" the assembly context from different sources,
and thus can restrain the overfitting.

\textbf{Effect of Symbolization.}
Similarly, we fixate the Siamese neural network (FCN feature extraction layers) of \sysname, 
and compare different symbolization policies on the icall dataset 
(Settings 1, 5 in Table~\ref{tab:ablation}).
As shown, strict symbolization has worse performance than loose symbolization. 
It confirms that the strict symbolization discards too much data-flow information, as discussed in Section~\ref{Design}. 
Additionally, the performance of strict symbolization degrades steeply (15.6\% \textit{F1}) from training to testing, which means that strict symbolization leads to worse over-fitting. 
In other words, strict symbolization leads to poor generalization performance. 
{\em Therefore, embedding with loose symbolization could better preserve data-flow information.}

\textbf{Effect of Feature Extraction Layers of the Siamese Neural Network.}
We have tested the performance of Siamese networks with different feature extraction layers
on the icall dataset (Settings 1, 6, 7, 8 in Table~\ref{tab:ablation}). 
The FCN we test has 3 hidden layers with 512 neurons.
The LSTM model has 512 neurons. 
The 1dCNN has 1 convolutional layer with 512 filters.
The TextCNN is adopted from \cite{kim2014convolutional}. 
We use ReLU as the activation function for these models.
As shown, Siamese networks with FCN layers have the best performance, 
achieving an \textit{F1} of 66.8\%. 
TextCNN layers perform slightly worse than FCN layers, with an \textit{F1} of 66.1\%.
1dCNN layers perform best on the training set but have the worst overfitting, 
leading to relatively poor performance on the testing set. 
The \textit{F1} drops 11.6\% from training to testing. 
LSTM layers have the worst performance. 
One explanation is that Recurrent Neural Networks such as LSTM 
usually take longer to converge due to the vanishing 
and exploding gradient problems \cite{bengio1994learning},
even if LSTM tried to ease gradient problems by introducing gates \cite{hochreiter1997lstm}.
Overall, we choose FCN layers as feature extraction layers.

\vspace{-0.1cm}
\subsubsection{Generalization across Compilers and Program Versions} \label{sssec:xeval}
\begin{figure}[!]
    \centering
    \subfigure[Cross-Compiler]{
    \includegraphics[width=3.7cm]{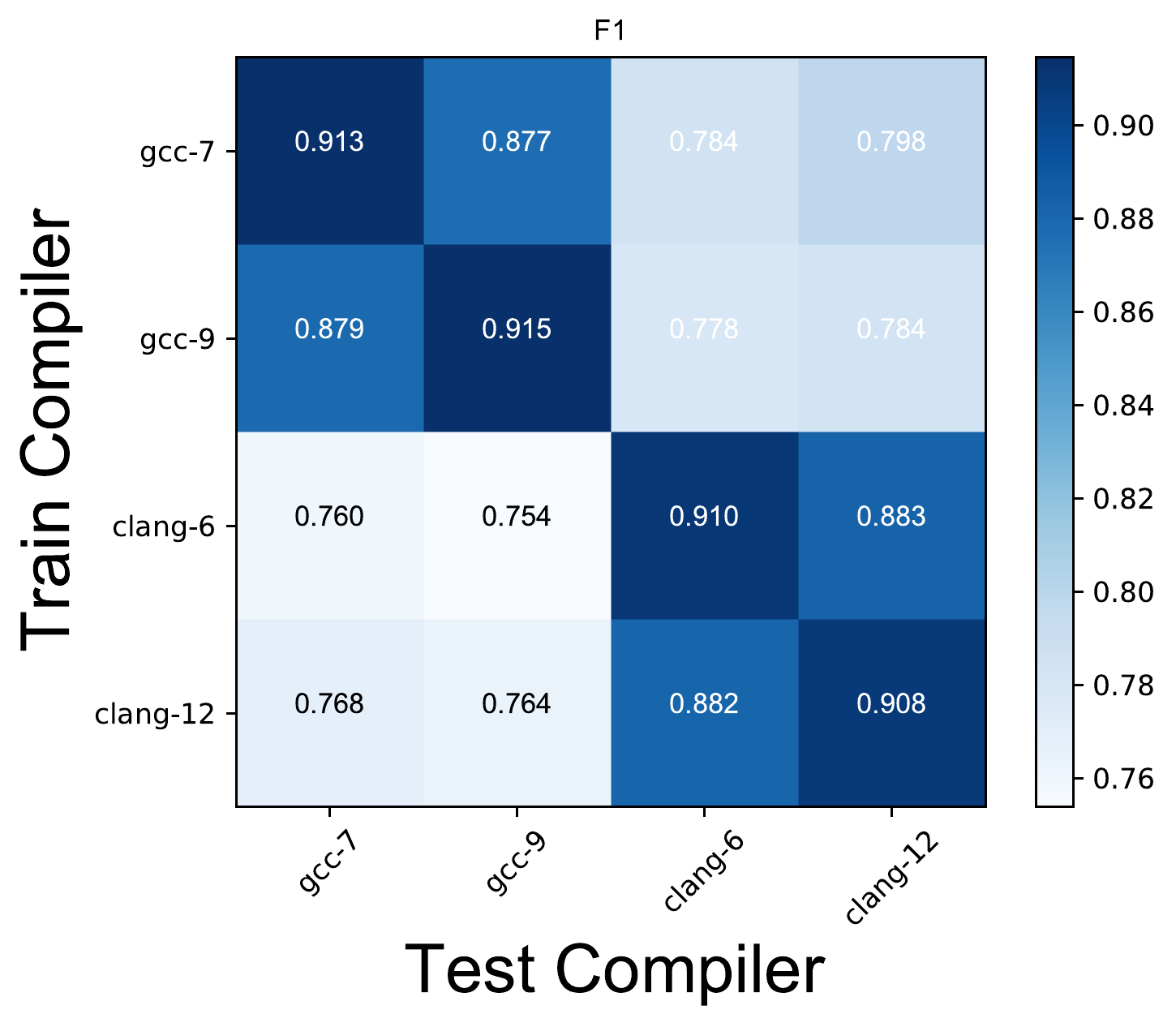}
    }
    \quad
    \subfigure[Cross-Version]{
    \includegraphics[width=3.7cm]{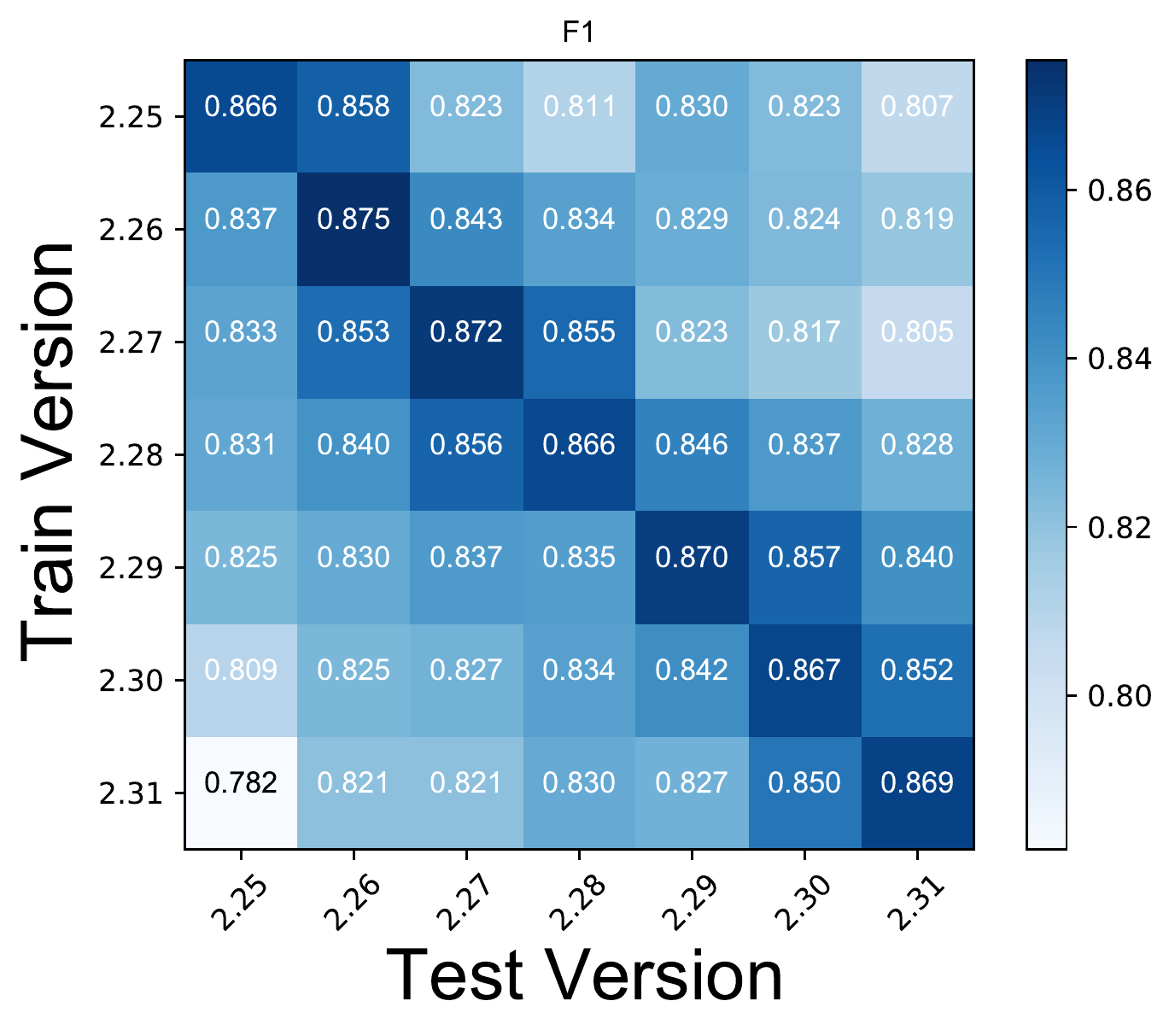}
    }
    \setlength{\abovecaptionskip}{-1mm}
    \caption{Generalization performance on GNU Binutils.}
    \label{fig:xcompiler_xversion}
    \vspace{-1mm}
\end{figure}

Apart from the generalization ability across binaries, we also 
evaluate the generalization ability across compilers and program versions.
The zero-shot learning results have shown that icall pairs share common patterns
with direct ones, and we thus believe they have common behavior in generalization.
Therefore, we perform \deleted{the generalization} experiments based on the large-scale dcall dataset.
Specifically, we build 7 versions (from 2.25 to 2.31)
of GNU Binutils with 4 compilers (gcc-7, gcc-9, clang-6, clang-12)
and further extract dcall pairs from them.
To evaluate the generalization ability across compilers,
we train the model on pairs from binaries compiled with one compiler (e.g. gcc-7)
and test on pairs from binaries compiled with another\deleted{compiler} (e.g. gcc-9).
Figure~\ref{fig:xcompiler_xversion}(a) shows the \textit{F1} of the cross-compiler setting. 
Data in the diagonal line indicates the upper limit of the model,
where training set and testing set are the same. 
In most difficult scenarios such as clang-12 vs gcc-7, 
whose generated assembly can have huge differences, 
the model can still achieve an \textit{F1} of 76\%, 
and in easier scenarios such as gcc-7 vs gcc-9, 
the model achieves a substantial performance with only a 2\%-4\% drop of \textit{F1}.
The model behaves likewise in the cross-version setting, as shown in Figure~\ref{fig:xcompiler_xversion}(b).
Across two most different versions 2.31 vs 2.25, to which 44 contributors have pushed over 5,000 commits~\cite{BinutilsDiff}, the model still achieves a 78\% \textit{F1}.
Thus \sysname has a substantial generalization performance in both
cross-compiler and cross-version settings.

\vspace{-0.1cm}
\subsubsection{Time Efficiency} \label{sssec:timeeval}
Suppose a binary has \textit{M} icallsites and \textit{N} candidate callees,
\sysname pair the callsites with each candidate callee and output a score for each input pair,
so the time complexity is \textit{O(MN)}. However, only address-taken functions are considered as possible candidates. And modern machine learning frameworks such as PyTorch provide batch inference, which takes advantage of scalable computation resources to generate many predictions at once. Suppose the batch\_size is \textit{B}, the time complexity will be \textit{O($\frac{MN}{B}$)}. Ideally, if the  RAM is sufficient to load all pairs, i.e. \textit{B=MN}, the model only needs to infer once.

After the one-time-effort pre-train, 
we measure the time consumption of key parts of \sysname with merely CPU.
It takes about 23s to fine-tune the doc2vec model and 2,407s to fine-tune the Siamese network.
After fine-tuning, on average, it takes about 0.0027s to perform slicing for a callsite-callee pair, 
0.0042s to embed a slice with the doc2vec model
and 0.0011s to infer one pair with the Siamese network.
For binaries in Table~\ref{tab:ICT},
it takes 4\textasciitilde30 seconds in total to analyze a binary with \sysname
and 6\textasciitilde45 seconds with TypeArmor.
However, as a pointer analysis, 
BPA needs more than 100 seconds to analyze small programs such as \texttt{lighttpd},
and more than 7 hours to analyze large programs like \texttt{gcc}.

In summary, we could draw the following conclusion: \textit{\sysname is more efficient and effective at recognizing icallees than state-of-the-art solutions such as BPA, TypeArmor as well as binary analysis tools.}

\vspace{-0.1cm}
\subsection{Applications of \sysname} \label{ssec:applications}
\vspace{-0.1cm}
\subsubsection{Promoting binary similarity detection}
With the final network trained and fine-tuned with pairs of all optimization levels,
we utilize \sysname to promote a fundamental task in binary similarity detection: binary diffing. 

The state-of-the-art solution DeepBinDiff~\cite{duan2020deepbindiff} 
leverages the program-wide control flow information to generate basic block embeddings. 
Specifically, it relies on an inter-procedural CFG (ICFG) generated by Angr, which is a combination of CGs and CFGs, 
to provide program-wide contextual information. 
Given two binaries, DeepBinDiff first generates an ICFG for each binary, 
merges them based on library functions, 
and runs the Text-associated DeepWalk (TADW) algorithm~\cite{yang2015tadw} to embed basic blocks. 
With generated embeddings, DeepBinDiff utilizes a k-hop greedy matching algorithm 
to match basic block pairs. 
In principle, if two icallsites in two binaries have similar callees, 
the two basic blocks they belong to should be similar too. 
Therefore, we can speculate that, \textit{with the CGs recovered by \sysname, 
DeepBinDiff would have better performance.}

Our experiments are performed on the same set of binaries used by DeepBinDiff, 
i.e., \texttt{printenv}, \texttt{md5sum}, \texttt{split}, \texttt{uniq}, \texttt{ls}, 
\texttt{who}, \texttt{cp}, \texttt{rmdir}, \texttt{yes}, \texttt{tty} 
from five versions of GNU Coreutils (v5.93, v6.4, v7.6, v8.1, v8.3)
with four optimization options (O0, O1, O2, O3). 
The binaries are compiled with the same compiler Clang, 
and we adopt the same metric used by DeepBinDiff, 
which is Precision, Recall, and F1-score of basic block matching. 
Parameters of DeepBinDiff are fixed to \textit{k=4}, \textit{threshold=0.6}, 
which are the optimal parameters according to their paper.
To eliminate the influence introduced by randomness in TADW, 
we repeat each experiment three times and calculate the average metrics.

We compare the performance of DeepBinDiff 
in diffing binaries across different versions and optimization levels, based on the original CGs and the CGs recovered by \sysname respectively. 
To further verify the usefulness of CGs recovered by \sysname, 
we also tested DeepBinDiff on crafted CGs that are generated by adding random edges between icallsites and potential callees.

\begin{table}[b]
\footnotesize
\setlength{\abovecaptionskip}{0cm}
\setlength{\belowcaptionskip}{0cm}
\caption{Cross-optimization-level binary diffing F1 scores of DeepBinDiff on the original CGs, on CGs with random edges, and on CGs recovered by \sysname.}
\label{tab:crossopt}
\centering
\begin{tabular}{c|ccc}
\toprule[1pt]
Optimization Levels         & DeepBinDiff & +Rand  & +\sysname          \\ \hline
O3 vs O2 & 89.0\%      & 85.3\% & \textbf{93.7\%} \\ 
O3 vs O1 & 69.7\%      & 67.8\% & \textbf{78.4\%} \\ 
O3 vs O0 & 10.8\%      & 9.3\%  & \textbf{25.6\%} \\ 
O2 vs O1 & 74.5\%      & 72.0\% & \textbf{92.1\%} \\ 
O2 vs O0 & 11.2\%      & 9.9\%  & \textbf{28.6\%} \\ 
O1 vs O0 & 13.7\%      & 12.8\% & \textbf{32.6\%} \\ \hline
Average  & 44.8\%      & 42.9\% & \textbf{58.5\%} \\ 
\bottomrule[1pt]
\end{tabular}
\end{table}

\textbf{Cross-optimization-level diffing.} 
Table~\ref{tab:crossopt} shows the F1-scores of cross-optimization-level diffing. 
We compile Coreutils-v7.6 and setup 6 experiments 
(O3 vs O2, O3 vs O1, O3 vs O0, O2 vs O1, O2 vs O0, O1 vs O0). 
As shown, compared to the original CGs, 
adding random edges would cause DeepBinDiff drop a 1.9\% F1-score (i.e., from 44.8\% to 42.9\%) on average, 
while adding edges recovered by \sysname would cause DeepBinDiff to increase the F1-score by 13.7\% (i.e., from 44.8\% to 58.5\%) on average. 
Detail statistics of the F1 scores of DeepBinDiff in different settings on different binaries are presented in Appendix~\ref{appendix:stat-bindiff}.

Note that, adding random edges decreases all settings' F1-scores, 
because it would significantly change the contexts of basic blocks that ought to be similar.
Whereas adding edges recovered by \sysname increases all settings' F1-scores, 
showing that precise CGs are useful for binary diffing and \sysname is effective at recovering CGs.

\begin{table}[t]
\setlength{\abovecaptionskip}{0cm}
\setlength{\belowcaptionskip}{0cm}
\caption{Cross-version Binary Diffing Results.}
\label{tab:crossversion}
\centering
\begin{tabular}{c|ccc}
\toprule[1pt]
Versions          & DeepBinDiff & +Rand    & +\sysname            \\ \hline
v5.93 vs v8.3 & 72.5\%      & 70.6\%   & \textbf{78.2\%}   \\
v6.4 vs v8.3  & 75.9\%      & 73.3\%   & \textbf{85.8\%}   \\
v7.6 vs v8.3  & 95.5\%      & 93.3\%   & \textbf{96.7\%}   \\
v8.1 vs v8.3  & 97.1\%      & 94.6\%   & \textbf{98.8\%}   \\ \hline
Average       & 85.3\%      & 83.0\%   & \textbf{89.9\%} \\ 
\bottomrule[1pt]
\end{tabular}
\end{table}

\textbf{Cross-version diffing.} 
Table~\ref{tab:crossversion} shows the F1-scores of cross-version diffing. 
We fix the Coreutils' optimization level to O1, and perform 4 experiments 
(v5.93 vs v8.3, v6.4 vs v8.3, v7.6 vs v8.3, v8.1 vs v8.3). 
Compared with DeepBinDiff, 
adding random edges leads to a 2.3\% F1-score decrease on average, 
while adding edges recovered by \sysname increases the F1-score by 4.6\% on average. 
Detailed statistics in different settings on different binaries are presented in Appendix~\ref{appendix:stat-bindiff}.
Consistent with the cross-optimization-level diffing results, 
we can see that, 
adding random edges decreases all settings' F1-scores and 
adding \sysname edges behaves in contrast.

Additionally, the evaluation shows that,
compared with cross-version diffing, cross-optimization-level diffing is more difficult, 
and larger increments appear in the cross-optimization-level settings involving the O0 level, 
i.e. O3-O0, O2-O0, O1-O0, compared with other settings. 
It indicates that optimization levels' effect is larger than versions', 
which is consistent with conclusions of DeepBinDiff and BINKIT~\cite{kim2020revisiting}. 
Thus we can obtain larger promotion in cross-optimization-level diffing by complementing the ICFG.

In summary, \sysname can improve the performance of DeepBinDiff by a large margin, 
especially in the cross-optimization-level diffing task. 

\vspace{-0.1cm}
\subsubsection{Promoting hybrid fuzzing}

We further apply \sysname to hybrid fuzzing.
Driller~\cite{stephens2016driller} is a hybrid fuzzer 
that augments the famous grey-box fuzzer AFL~\cite{AFL} with symbolic execution
\deleted{.Driller will invoke its symbolic execution engine} when AFL gets stuck.
Specifically, driller takes all untraced paths\deleted{which exist} in AFL's queue 
and looks for basic block transitions AFL failed to find satisfying inputs for. 
Driller will then use Angr to solve inputs for these\deleted{basic block} transitions and pass them to AFL. 
However, driller does not monitor transitions invoked by icalls,
and thus \textit{we could speculate that augmenting driller with the CGs recovered by \sysname
can help driller cover more paths, i.e., improve the code coverage.}
\added{
We have modified driller to solve symbolic constraints to generate testcases when it hits an icall.
}

Our experiments are performed on the same binaries used by driller, 
i.e., the DARPA CGC chanllenges~\cite{darpa2014cgc}.
We choose all 8 challenges that involve icall in the code,
and fuzz the binaries for 24 hours,
Experiments are repeated 3 times and we calculate the average number of results.
We compare the number of triggered paths\deleted{and unique crashes} of each challenge between
the vanilla driller and driller with icall resolving based on the CGs recovered by \sysname.
Analogically, we also include a driller with icall resolving based on the CGs 
with added random edges.

As shown in Table~\ref{tab:hybridfuzzing}, on average, adding random edges to CGs decreases the number 
of paths by 8\deleted{and number of unique crashes by 2}, while adding edges recovered by
\sysname can increase the numbers by over 50\%,
because adding random edges could misguide the symbolic execution engine to solve
unreachable edges. 
\deleted{Whereas adding edges recovered by \sysname can increase all challenges'
code coverage and crash numbers, demonstrating the effectiveness of \sysname.}
\added{Note that the fuzzer may spend more time solving symbolic constraints introduced by icalls 
than conditional branches, known as the exploration-exploitation trade-off problem, 
as shown in \texttt{NRFIN\_00026} and \texttt{KPRCA\_00003}. 
But overall adding edges recovered by \sysname can increase the code coverage, 
demonstrating the effectiveness of \sysname.
We additionally examine the crashes found by the fuzzers, 
and results show that adding icall edges can also benefit slightly.
For example}\deleted{Specifically}, 
on the \texttt{KPRCA\_00017} challenge, vanilla driller and driller+Rand failed to
trigger crash within 24 hours \deleted{, but}\added{while} driller+\sysname can.\deleted{trigger crashes 23 times.}

\begin{table}[]
\centering
\setlength{\abovecaptionskip}{0cm}
\setlength{\belowcaptionskip}{0cm}
\caption{Hybrid Fuzzing Results.}
\label{tab:hybridfuzzing}
\resizebox{.48\textwidth}{!}{
\begin{tabular}{c|ccc|ccc}
\toprule[1pt]
\multirow{2}{*}{Challenge}    & \multicolumn{3}{c|}{\# Paths} & \multicolumn{3}{c}{Found Crash?} \\ \cline{2-7} 
                              & Driller   & +Rand  & +\sysname  & Driller   & +Rand   & +\sysname   \\ \hline
NRFIN\_00026                    & 26        & 25      & 20     & \deleted{0}\ding{56}         & \deleted{0}\ding{56}        & \deleted{0}\ding{56}       \\
LUNGE\_00002                    & 39        & 37      & 120    & \deleted{9}\added{\checkmark}         & \deleted{7}\added{\checkmark}        & \deleted{10}\added{\checkmark}      \\
YAN01\_00007                    & 45        & 45      & 125    & \deleted{0}\ding{56}         & \deleted{0}\ding{56}        & \deleted{0}\ding{56}       \\
NRFIN\_00074                    & 412       & 404     & 489    & \deleted{76}\added{\checkmark}        & \deleted{68}\added{\checkmark}       & \deleted{95}\added{\checkmark}      \\
KPRCA\_00017                    & 246       & 221     & 283    & \deleted{0}\ding{56}         & \deleted{0}\ding{56}        & \deleted{23}\added{\checkmark}      \\
KPRCA\_00003                    & 11        & 10      & 9      & \deleted{1}\added{\checkmark}         & \deleted{1}\added{\checkmark}        & \deleted{1}\added{\checkmark}       \\
KPRCA\_00060                    & 140       & 113     & 394    & \deleted{10}\added{\checkmark}        & \deleted{3}\added{\checkmark}        & \deleted{17}\added{\checkmark}      \\
NRFIN\_00076                    & 45        & 42      & 47     & \deleted{0}\ding{56}         & \deleted{0}\ding{56}        & \deleted{0}\ding{56}       \\ \hline
Average                         & 120.5     & 112.1   & 185.9  & \deleted{12.0}\added{-}      & \deleted{9.9}\added{-}      & \deleted{18.3}\added{-}    \\
\bottomrule[1pt]
\end{tabular}
}
\vspace{-0.3cm}
\end{table}

In summary, we could draw the following conclusion:
\textit{\sysname can promote \deleted{CG-based tasks such as binary similarity detection and hybrid fuzzing}\added{binary similarity detection and improve the code coverage in hybrid fuzzing.}}

\vspace{-0.1cm}
\subsection{Interpretability of \sysname} \label{ssec:interpre}
\vspace{-0.1cm}
To examine whether \sysname has learned interpretable knowledge, 
we visualize the embedding model as well as the weights of the Siamese neural network.
\subsubsection{Embedding Model}
\begin{figure}[t]
    \centering
    \includegraphics[width =.49\textwidth]{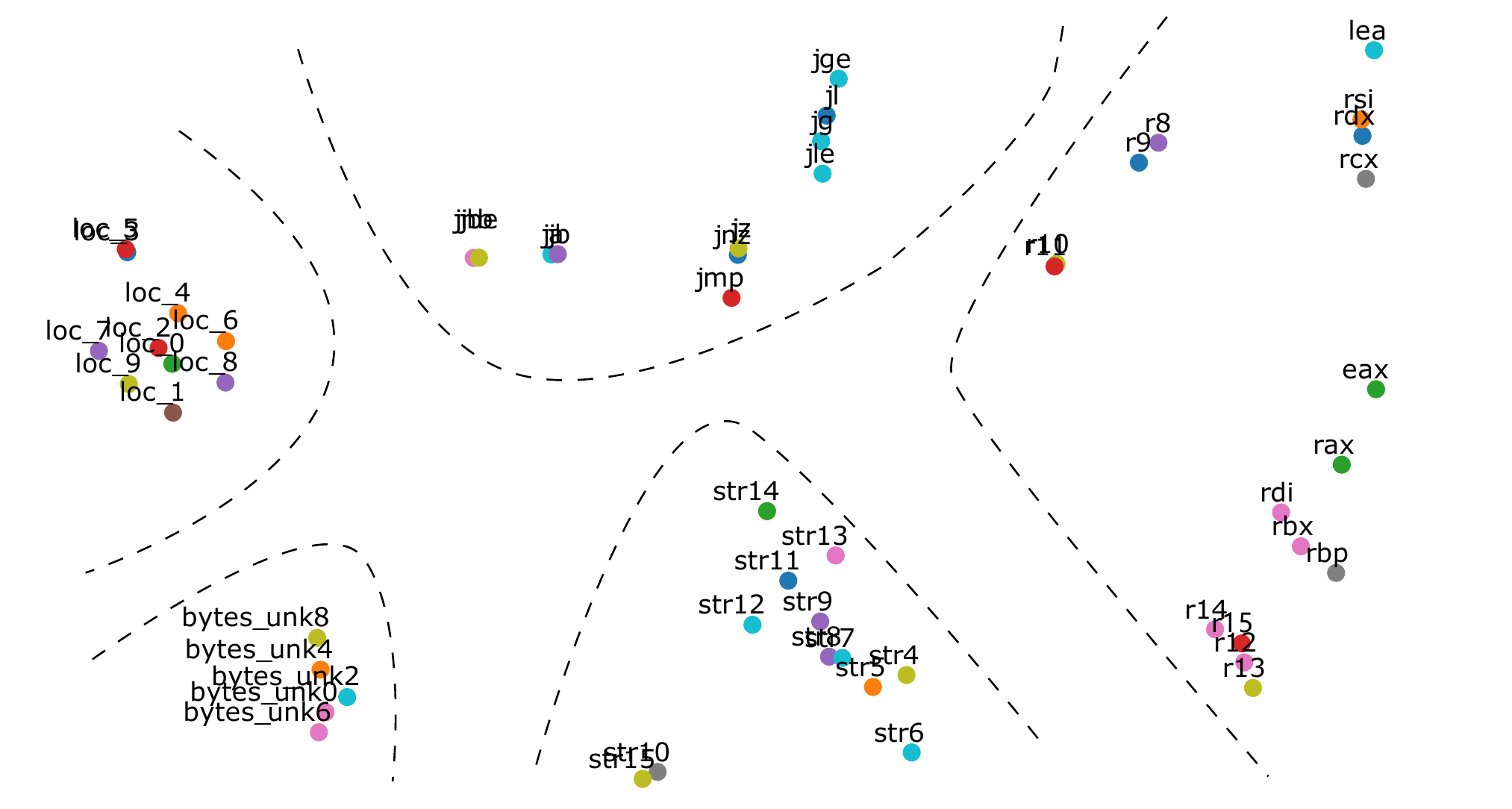}
    \caption{T-SNE visualization of tokens in doc2vec}
    \label{fig:tsne}
     \vspace{-0.4cm}
\end{figure}

\quad
We use T-SNE~\cite{maaten2008visualizing} to project high-dimensional vectors 
to a 2D space to examine whether the embedding model could group semantically-close tokens together.
There are 3,330 tokens after Loose symbolization. The smaller the distance between tokens, 
the more similar their semantic features are. For example, token \texttt{jb} and \texttt{jnb} are both instructions related to conditional jump, 
so they are clustered together in Figure~\ref{fig:tsne}. 
Therefore, word vectors generated by the doc2vec model can well capture semantic features of tokens in assembly instructions.

\subsubsection{Siamese network}\quad
We utilize the saliency map to interpret the network
to deduce the sensitivity of output regarding input vectors.
First, we compute partial derivatives for input pairs. 
Given a callsite or callee slice (after vectorization) $x \in R^{l \times d}$, $l$ is the length of the slice, and $d$ is the dimension of a token's embedding.
$f(x)$ is the output of the Siamese network. The partial derivatives is given by:
\vspace{-0.1cm}
\[
\nabla_{x}f(x)=\frac{\partial f}{\partial x}=[\frac{\partial f}{\partial x_{i,j}}]_{i \in 1...l, j \in 1...d}
\]
\vspace{-0.1cm}

This partial derivative consists of gradients of each input token. 
To measure the sensitivity of each token, we further compute the magnitude of gradient. The saliency map $S(x)$ is defined as:
\vspace{-0.1cm}
\[
S(x)[i]=\sqrt{(\frac{\partial f}{\partial x_{i,1}})^{2}+(\frac{\partial f}{\partial x_{i,2}})^{2}+...+(\frac{\partial f}{\partial x_{i,d}})^{2}}
\]
\vspace{-0.1cm}

\begin{figure}[t]
    \centering
    \includegraphics[width = .32\textwidth]{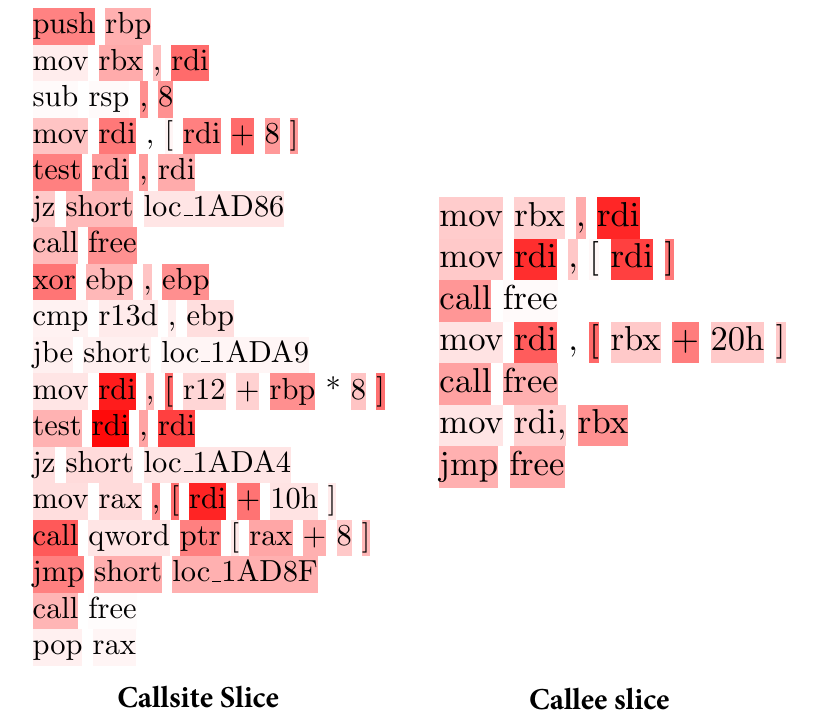}
    \caption{Saliency map of the pair from \texttt{lighttpd}.}
    \label{fig:heatmap}
     \vspace{-0.5cm}
\end{figure}

With the saliency map to interpret \sysname,
we present a case study of a pair from {\tt lighttpd} on which \sysname surpasses TypeArmor. 
With the help of debug info, 
we could map the assembly pair to source code:
the callsite is \texttt{a->data[i]->fn->free(a->data[i])} in function \texttt{array\_free\_data},
and the callee is function \texttt{void array\_data\_string\_free(ptr *p)}.
However, TypeArmor wrongly reports the callee as "non-void" function, 
and thus could lead to type-matching mistakes. 
\sysname predicted the pair as "match", and the saliency map is shown in Figure~\ref{fig:heatmap}.
In the saliency map, a token with darker color means a larger $S(x)[i]$,
i.e. a greater contribution to model decision, according to the definition of saliency map. 
Thus in the slices of the callsite and callee, 
the most important tokens are all related to the argument register \texttt{rdi},
and meanwhile tokens concerning the return value register \texttt{rax} has little contribution.
It demonstrates that the network indeed can capture important features of the calling convention. 
In other words, the network has learned patterns consistent with domain knowledge.

In summary, 
we could draw the following conclusion:
{\em The embedding model reasonably represents tokens in a high-dimensional space,
and the Siamese neural network can learn patterns consistent with domain knowledge.}

\vspace{-0.2cm}
\section{Discussion and Limitations} \label{Discussion}
\vspace{-0.1cm}
\added{
\textbf{Cross-optimization-level evaluation of \sysname.}
Cross-optimization-level callsite-callee matching, e.g. training with GCC-O0 pairs only and testing with GCC-O3 pairs, is not common in production environments. Instead, \sysname trains \textit{one unified} model with pairs of functions that are compiled with all optimization levels before deployment to answer users' callsite-callee matching questions. 
Nevertheless, we have compared the performance of this unified model with the performance of multiple models for  individual optimization levels, i.e., each model is only trained with pairs of functions compiled with the same optimization level,
and results show that our unified model has very close performance (less than 1\% F1). 
Thus we use the unified model for downstream applications.
}

\textbf{Mechanism of neural networks.}
Although we have used T-SNE to visualize the distribution of token embeddings 
and calculated the saliency map of the Siamese network, 
\sysname is designed to provide a {\em reference for}, 
rather than teaching human experts to analyze binaries,
because the robustness of interpretation of neural networks has not been theoretically proved~\cite{ghorbani2019interpretation},
and currently there is no standard method to interpret DNNs for binary analysis.

\textbf{Indirect jumps.} 
Currently, \sysname only handles icalls and does not support indirect jumps. 
In general, indirect jumps are used for \texttt{switch} statements or tail calls.
For the former, their targets can be recovered from the associated jump table generated by compilers~\cite{lu2019does}. 
For the latter, they are almost the same as icalls.
Our solution could be extended to support them in the same way, i.e., slicing, preprocessing, embedding and matching with a Siamese network.

\textbf{Variadic functions.}
Type-based solutions\deleted{, whether at binary-level or source-level,} cannot well support variadic functions,
i.e. functions with a variable number of arguments. 
While \sysname matches callsites with callees by apprehending their contexts and has no requests on the arguments. 
As long as the instructions concerned with arguments are all kept in the context, 
the network can extract features automatically from the context.

\textbf{Applicability to programs with other calling conventions or in other architectures or obfuscated.}
The software ecosystem has various calling conventions and architectures.
\deleted{Other calling conventions differ from the calling convention of the System V AMD64 ABI. }
For example, for 32-bit programs using the x86 cdecl calling convention, 
function arguments are passed via the stack.
Another example is that, smart contracts written in Solidity run in a stack-based virtual machine.
To apply \sysname to \deleted{32-bit x86 programs}\added{programs with other calling conventions}, 
one \deleted{can}\added{needs to} adjust the current policies of slicing and symbolization.
In the same way can one apply \sysname to programs in other architectures or obfuscated ones. 
Overall, the idea of comprehending contexts of callsites and callees and 
matching them in a question-answering way 
is theoretically reasonable for all programs.
We leave it as future work.

\textbf{Applicability to tasks that require a 100\% recall.}
Tasks such as Control-flow integrity (CFI) and binary rewriting usually require a 100\% recall
to avoid compatibility issues caused by false negatives.
However, due to the random nature of neural networks, one cannot ensure neural networks achieve a 100\% recall, therefore to apply \sysname to those tasks, additional efforts are required to eliminate false negatives.
Actually, even TypeArmor can have false negatives as well~\cite{kim2021icall},
and BPA achieves a 100\% recall on top of binary profiling.
Except for binary profiling, one can 
ease the false-negative problem by increasing the matching threshold, 
while introducing more false positives.

\textbf{Working on assembly rather than IR.}
Lifting binaries to IR actually relies on indirect control-flow resolution~\cite{altinay2020binrec}. 
Besides, existing binary lifting tools can generate redundant or even incorrect IR \cite{kim2017meandiff}.
Therefore, we believe that lifting binaries to IR may lead to more information loss, enlarging the difficulty for neural networks to comprehend the context.

\vspace{-0.2cm}
\section{Conclusion} \label{Conclusion}
\vspace{-0.2cm}
In this paper, we present \sysname, a transfer- and contrastive-learning approach that effectively recognizes icallees at the binary level. 
By slicing the contexts of callsites and callees,
\sysname trains an assembly-centric doc2vec model 
and a Siamese neural network to match callsites with callees.
Evaluation results show that,
\sysname can recognize icallees with high precision and recall, 
and can recover call graphs to promote downstream applications, 
e.g., binary code similarity detection and hybrid fuzzing. 
By interpreting the embedding model and the Siamese neural network, 
we demonstrate that \sysname learns knowledge similar to human experts, and thus can apprehend the assembly language to some extent. 
Therefore, we believe that transfer-learning approaches are promising for binary program analysis tasks.

\vspace{-0.1cm}
\section*{Acknowledgment}
\vspace{-0.1cm}
We thank the anonymous reviewers and our shepherd for their insightful feedback,
especially the suggestion of transfer learning.
This work was supported 
in part by the National Key Research and Development Program of China (2021YFB2701000, 2021YFB3101200), National Natural Science Foundation of China (61972224, 62272265, U1836213), and Beijing National Research Center for Information Science and Technology (BNRist) under Grant BNR2022RC01006.
Any findings are those of the authors and do not necessarily reflect the views of our sponsors.

{
\footnotesize
\bibliographystyle{IEEEtran}
\bibliography{ref}
}

\appendix
\subsection{Callsite-callee pair collection} \label{appendix:datacollection}

\textbf{User-mode binaries.} 
For user-mode binaries, 
we first turn off the Address Space Layout Randomization (ASLR) for convenience, 
then we have tried the following methods: 

\begin{itemize}[leftmargin=.32cm,noitemsep,topsep=2pt]
    
    \item LLVM. We instrument all indirect callsites by an LLVM machine pass. 
    When compiling binaries, this pass identifies all indirect call instructions, and inserts a one-byte int3 instruction before them. 
    We then write a debugger script to automatically catch breakpoints caused by this instruction and record runtime information, including callsite addresses, the callee addresses, and virtual memory maps of the binaries
    (to recognize addresses resided in shared libraries).
    \item Fuzzing \& Intel Processor Tracing (PT).
    We first use coverage-guided fuzzers such as American Fuzzy Loop (AFL) \cite{AFL} to get inputs that can cover as much code as possible. 
    Then run the program with these inputs, and use Intel PT \cite{PT} to record execution traces. 
    Finally, with the libipt \cite{libipt} decoder library, we extract indirect call instructions from the trace, take their next instructions as targets and make pairs.

\end{itemize}

\begin{figure}[ht]
    \centering
    \setlength{\abovecaptionskip}{0cm}
    \includegraphics[width = .7\linewidth]{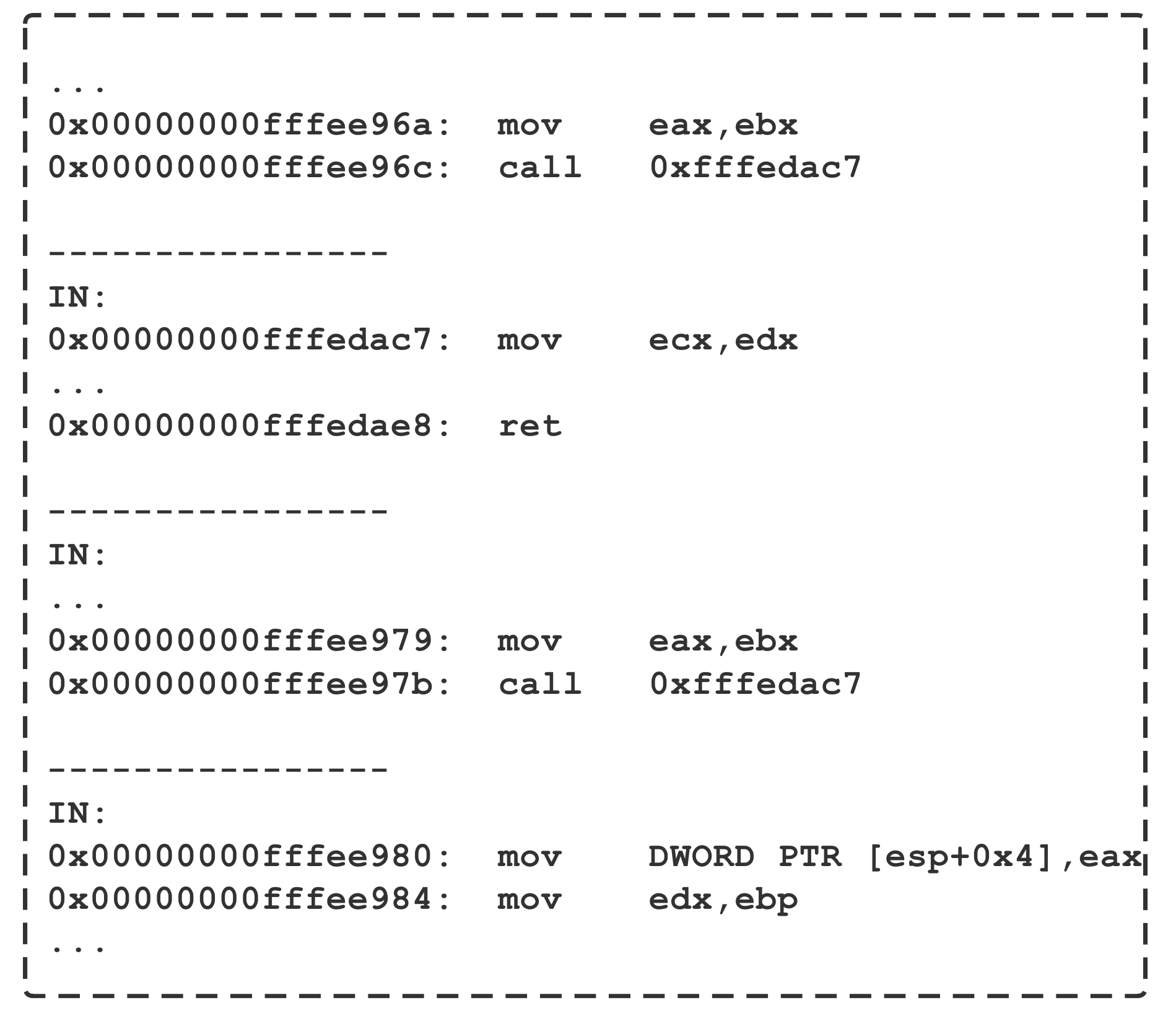}
    \caption{Logging optimization of PANDA. Function call \texttt{call 0xfffedac7} is continuously invoked twice at address \texttt{0x00000000fffee96c} and address \texttt{0x00000000fffee97b}, but the function body (instructions) is only recorded once.}
    \label{fig:panda}
\end{figure}

\textbf{The Linux kernel.} 
Likewise, we turn off Kernel Address Space Layout Randomization (KASLR) when compiling the kernel for the convenience of implementation. If KASLR is on, addresses recorded during runtime are complicated to be mapped back to the static addresses in the binary.
Afterward, the kernel is emulated in an open-source record and replay platform PANDA~\cite{dolan2013panda},
which is built upon the QEMU~\cite{bellard2005qemu} whole system emulator. 
We enable the \texttt{"-d in\_asm"} option of PANDA to log the target assembly code and instruction addresses.

Kernel traces are stored in a log file, from which we can extract the addresses of callsite-callee pairs.
Usually, the next instruction of a callsite should be the target callee,
however, there are two challenges in parsing the kernel trace log: 

\begin{itemize}[leftmargin=.32cm,noitemsep,topsep=2pt]
    \item Hardware interrupt. When a hardware interrupt is encountered right after an indirect call, we do not record the current pair, 
    since we have no knowledge of hardware interrupts.
    \item Logging optimization of PANDA. As shown in Figure~\ref{fig:panda}, when a function is invoked multiple times, 
    PANDA may log function body texts only once in the trace. 
    Hence we check indirect calls which are continuously invoked.
    To avoid false callees, we only record the target of the first indirect call (i.e. address of the first callsite's next instruction).
\end{itemize}

\textbf{Rational behind the dynamic analysis to collect ground truths.}
Recall that data as ground truths should all be true positives. 
And an icall that can be invoked during runtime without violating sanitizers 
is always legitimate and thus dynamically collected icall pairs are all true positives. 
Although potential legitimate pairs might be missed during dynamic analysis, the collected ground truths are 100\% accurate. 
Besides, although dynamically-collected icall pairs can be easy-to-trigger, 
it is orthogonal to the callsite-callee matching
because the complexity of a callsite's control-flow constraints does not influence
the validity of its callees.

\subsection{Detail statistics of DeepBinDiff} \label{appendix:stat-bindiff}
The performance of DeepBinDiff depends on the call graphs it can get. 
In this section, we present the detailed F1 scores of DeepBinDiff on different binaries in different settings.

In the cross-optimization-level binary diffing setting, 
the F1 scores of DeepBinDiff based on the original CGs, CGs with random edges and CGs with edges recovered by \sysname are shown in Table~\ref{tab:vanillaoptdiff}, Table~\ref{tab:randoptdiff} and Table~\ref{tab:icalloptdiff} respectively.

In the cross-version binary diffing setting, 
the F1 scores of DeepBinDiff based on the original CGs, CGs with random edges and CGs with edges recovered by \sysname are shown in Table~\ref{tab:vanillaverdiff}, Table~\ref{tab:randverdiff} and Table~\ref{tab:icallverdiff} respectively.

\begin{table*}[ht]
    \footnotesize
\setlength{\abovecaptionskip}{0cm}
\caption{Cross-optimization-level binary diffing F1 scores  of DeepBinDiff, based on \textbf{original} CGs.}
\label{tab:vanillaoptdiff}
\centering
\begin{tabular}{c|cccccccccc|c}
\toprule[1pt]
Optimization Levels & printenv & md5sum & split  & uniq   & ls     & who    & cp     & rmdir  & yes    & tty    & Average \\ \hline
O3 vs O2  &  87.8\%  &  89.4\%  &  91.4\%  &  87.8\%  &  84.8\%  &  91.9\%  &  92.1\%  &  90.7\%  &  87.6\%  &  86.8\%  &  89.0\%   \\
O3 vs O1  &  72.7\%  &  72.9\%  &  75.0\%  &  69.8\%  &  60.5\%  &  65.8\%  &  72.4\%  &  64.6\%  &  72.0\%  &  71.6\%  &  69.7\%   \\
O3 vs O0  &  9.0\%  &  13.2\%  &  10.6\%  &  14.0\%  &  8.0\%  &  11.0\%  &  11.8\%  &  8.0\%  &  11.4\%  &  10.6\%  &  10.8\%   \\
O2 vs O1  &  78.4\%  &  78.1\%  &  79.4\%  &  75.7\%  &  67.8\%  &  68.6\%  &  74.8\%  &  66.9\%  &  77.0\%  &  77.8\%  &  74.5\%   \\
O2 vs O0  &  12.9\%  &  11.5\%  &  11.6\%  &  14.7\%  &  8.6\%  &  9.2\%  &  14.0\%  &  7.4\%  &  11.4\%  &  10.3\%  &  11.2\%   \\
O1 vs O0  &  13.4\%  &  13.7\%  &  14.2\%  &  15.8\%  &  9.8\%  &  14.2\%  &  14.1\%  &  10.0\%  &  15.8\%  &  16.4\%  &  13.7\%   \\ \hline
Average   &  45.7\%  &  46.5\%  &  47.0\%  &  46.3\%  &  39.9\%  &  43.5\%  &  46.5\%  &  41.3\%  &  45.9\%  &  45.6\%  &  44.8\%   \\
\bottomrule[1pt]
\end{tabular}
\end{table*}

\begin{table*}[ht]
    \footnotesize
\setlength{\abovecaptionskip}{0cm}
\caption{Cross-optimization-level binary diffing F1 scores  of DeepBinDiff, based on CGs instrumented  with \textbf{random edges}.}
\label{tab:randoptdiff}
\centering
\begin{tabular}{c|cccccccccc|c}
\toprule[1pt]
Optimization Levels & printenv & md5sum & split  & uniq   & ls     & who    & cp     & rmdir  & yes    & tty    & Average \\ \hline
O3 vs O2  &  83.8\%  &  86.5\%  &  87.1\%  &  85.8\%  &  81.4\%  &  85.9\%  &  87.9\%  &  85.4\%  &  84.1\%  &  85.4\%  &  85.3\%   \\
O3 vs O1  &  69.1\%  &  71.7\%  &  70.8\%  &  70.1\%  &  59.1\%  &  64.7\%  &  67.6\%  &  62.6\%  &  71.2\%  &  71.5\%  &  67.8\%   \\
O3 vs O0  &  8.1\%  &  10.2\%  &  8.8\%  &  9.5\%  &  7.8\%  &  8.8\%  &  11.8\%  &  7.0\%  &  11.8\%  &  8.7\%  &  9.2\%   \\
O2 vs O1  &  75.4\%  &  74.1\%  &  77.8\%  &  73.9\%  &  66.3\%  &  66.2\%  &  69.4\%  &  67.1\%  &  75.5\%  &  74.6\%  &  72.0\%   \\
O2 vs O0  &  10.9\%  &  11.1\%  &  9.4\%  &  12.5\%  &  8.3\%  &  8.2\%  &  10.3\%  &  7.0\%  &  10.3\%  &  11.4\%  &  9.9\%   \\
O1 vs O0  &  12.1\%  &  14.1\%  &  10.3\%  &  15.6\%  &  9.6\%  &  14.7\%  &  11.1\%  &  8.8\%  &  16.7\%  &  15.3\%  &  12.8\%   \\ \hline
Average   &  43.2\%  &  44.6\%  &  44.0\%  &  44.6\%  &  38.8\%  &  41.4\%  &  43.0\%  &  39.6\%  &  44.9\%  &  44.5\%  &  42.9\%   \\
\bottomrule[1pt]
\end{tabular}
\end{table*}

\begin{table*}[ht]
\footnotesize
\setlength{\abovecaptionskip}{0cm}
\setlength{\belowcaptionskip}{0cm}
\caption{Cross-optimization-level binary diffing F1 scores  of DeepBinDiff, based on CGs recovered by \textbf{\sysname}.}
\label{tab:icalloptdiff}
\centering
\begin{tabular}{c|cccccccccc|c}
\toprule[1pt]
Optimization Levels & printenv & md5sum & split  & uniq   & ls     & who    & cp     & rmdir  & yes    & tty    & Average \\ \hline
O3 vs O2    & 89.7\%    & 96.5\%  & 99.0\%  & 90.4\%  & 93.0\%  & 98.1\%  & 96.1\%  & 95.3\%  & 89.4\%  & 89.6\%  & 93.71\%  \\
O3 vs O1    & 76.4\%    & 76.0\%  & 74.5\%  & 76.2\%  & 87.5\%  & 81.5\%  & 76.7\%  & 81.6\%  & 78.0\%  & 75.4\%  & 78.38\%  \\
O3 vs O0    & 27.7\%    & 30.1\%  & 25.6\%  & 27.6\%  & 18.7\%  & 23.9\%  & 28.7\%  & 18.5\%  & 27.1\%  & 27.9\%  & 25.58\%  \\
O2 vs O1    & 87.6\%    & 93.6\%  & 92.2\%  & 92.7\%  & 93.6\%  & 94.3\%  & 95.7\%  & 97.6\%  & 87.3\%  & 86.1\%  & 92.07\%  \\
O2 vs O0    & 26.8\%    & 34.0\%  & 28.3\%  & 36.3\%  & 20.4\%  & 26.6\%  & 30.4\%  & 15.7\%  & 32.7\%  & 34.5\%  & 28.57\%  \\
O1 vs O0    & 36.7\%    & 33.3\%  & 31.4\%  & 37.0\%  & 26.1\%  & 32.3\%  & 32.7\%  & 24.7\%  & 35.8\%  & 35.7\%  & 32.57\%  \\ \hline
Average     & 57.5\%    & 60.6\%  & 58.5\%  & 60.0\%  & 56.6\%  & 59.5\%  & 60.1\% & 55.6\%   & 58.4\%  & 58.2\%  & 58.5\%   \\
\bottomrule[1pt]
\end{tabular}
\end{table*}

\begin{table*}[ht]
    \footnotesize
\setlength{\abovecaptionskip}{0cm}
\caption{Cross-version binary diffing F1 scores of DeepBinDiff, based on \textbf{original} CGs.}
\label{tab:vanillaverdiff}
\centering
\begin{tabular}{c|cccccccccc|c}
\toprule[1pt]
Versions          & printenv & md5sum & split  & uniq   & ls     & who    & cp     & rmdir  & yes    & tty    & Average \\ \hline
v5.93 vs v8.3      &  61.7\%  &  68.0\%  &  74.3\%  &  79.5\%  &  76.5\%  &  84.5\%  &  75.5\%  &  67.0\%  &  68.2\%  &  70.0\%  &  72.5\%   \\
v6.4 vs v8.3       &  67.8\%  &  77.2\%  &  79.7\%  &  82.2\%  &  80.5\%  &  87.3\%  &  76.2\%  &  69.5\%  &  67.4\%  &  71.4\%  &  75.9\%   \\
v7.6 vs v8.3       &  92.5\%  &  94.0\%  &  97.0\%  &  97.5\%  &  94.5\%  &  98.6\%  &  93.7\%  &  96.9\%  &  94.7\%  &  95.9\%  &  95.5\%   \\
v8.1 vs v8.3      &  97.9\%  &  97.9\%  &  97.3\%  &  98.4\%  &  95.1\%  &  96.7\%  &  95.5\%  &  97.6\%  &  97.7\%  &  97.2\%  &  97.1\%   \\ \hline
Average        &  80.0\%  &  84.3\%  &  87.1\%  &  89.4\%  &  86.7\%  &  91.8\%  &  85.2\%  &  82.8\%  &  82.0\%  &  83.6\%  &  85.3\%   \\
\bottomrule[1pt]
\end{tabular}
\end{table*}

\begin{table*}[ht]
    \footnotesize
\setlength{\abovecaptionskip}{0cm}
\caption{Cross-version binary diffing F1 scores  of DeepBinDiff, based on CGs instrumented with \textbf{random edges}.}
\label{tab:randverdiff}
\centering
\begin{tabular}{c|cccccccccc|c}
\toprule[1pt]
Versions          & printenv & md5sum & split  & uniq   & ls     & who    & cp     & rmdir  & yes    & tty    & Average \\ \hline
v5.93 vs v8.3  &  59.6\%  &  68.9\%  &  75.3\%  &  79.1\%  &  71.8\%  &  82.5\%  &  74.0\%  &  59.7\%  &  65.1\%  &  69.8\%  &  70.6\%   \\
v6.4 vs v8.3   &  65.5\%  &  71.2\%  &  72.5\%  &  83.5\%  &  78.9\%  &  81.7\%  &  73.2\%  &  70.8\%  &  66.7\%  &  69.2\%  &  73.3\%   \\
v7.6 vs v8.3   &  91.9\%  &  92.8\%  &  92.7\%  &  96.4\%  &  92.2\%  &  96.7\%  &  90.0\%  &  93.4\%  &  94.1\%  &  93.1\%  &  93.3\%   \\
v8.1 vs v8.3   &  97.0\%  &  97.3\%  &  92.6\%  &  97.6\%  &  93.8\%  &  92.0\%  &  92.3\%  &  93.2\%  &  95.8\%  &  94.8\%  &  94.6\%   \\ \hline
Average        &  78.5\%  &  82.5\%  &  83.3\%  &  89.1\%  &  84.2\%  &  88.2\%  &  82.4\%  &  79.3\%  &  80.4\%  &  81.7\%  &  83.0\%   \\
\bottomrule[1pt]
\end{tabular}
\end{table*}

\begin{table*}[ht]
    \footnotesize
\setlength{\abovecaptionskip}{0cm}
\caption{Cross-version binary diffing F1 scores  of DeepBinDiff, based on CGs recovered by \textbf{\sysname}.}
\label{tab:icallverdiff}
\centering
\begin{tabular}{c|cccccccccc|c}
\toprule[1pt]
Versions          & printenv & md5sum & split  & uniq   & ls     & who    & cp     & rmdir  & yes    & tty    & Average \\ \hline
v5.93 vs v8.3  &  65.6\%  &  79.7\%  &  81.4\%  &  83.5\%  &  87.4\%  &  84.8\%  &  80.7\%  &  70.7\%  &  73.7\%  &  74.7\%  &  78.2\%   \\
v6.4 vs v8.3   &  79.5\%  &  84.2\%  &  88.7\%  &  90.4\%  &  89.1\%  &  93.1\%  &  85.0\%  &  83.4\%  &  81.2\%  &  83.0\%  &  85.8\%   \\
v7.6 vs v8.3   &  93.5\%  &  95.4\%  &  98.6\%  &  96.0\%  &  99.0\%  &  98.0\%  &  97.7\%  &  95.3\%  &  97.0\%  &  96.3\%  &  96.7\%   \\
v8.1 vs v8.3   &  98.7\%  &  98.7\%  &  99.1\%  &  98.3\%  &  99.0\%  &  99.1\%  &  99.5\%  &  98.0\%  &  99.0\%  &  98.9\%  &  98.8\%   \\ \hline
Average        &  84.3\%  &  89.5\%  &  92.0\%  &  92.0\%  &  93.6\%  &  93.8\%  &  90.7\%  &  86.9\%  &  87.7\%  &  88.2\%  &  89.9\%   \\
\bottomrule[1pt]
\end{tabular}
\end{table*}

\clearpage

\subsection{\added{Impact of embedding techniques}} \label{appendix:embedding}

\added{To evaluate the generalization ability of different embedding techniques, 
we report the zero-shot performance,
i.e., precision, recall and F1 scores when training with dcall pairs and testing with icall pairs,
of 5 common embedding methods: Instruction2Vec~\cite{lee2017instruction2vec}, word2vec, PalmTree, Asm2Vec, and doc2vec.
Since Instruction2Vec and word2vec cannot generate embeddings for instruction sequences directly,
we have averaged instruction/token embeddings to obtain the sequence embedding.

As shown in Table~\ref{tab:embedding}, 
word2vec has the worst performance because it does not consider the internal structures of instructions;
Instruction2Vec achieves an acceptable performance for its fine-grained instruction embedding model.
Asm2Vec~\cite{ding2019asm2vec} performs better than Instruction2Vec but worse than PalmTree and doc2vec
becasue it generates instruction sequences by random walk, 
which may lead to illegitimate control-flow sequences and cannot guarantee code coverage.
F1-scores of PalmTree and doc2vec are close, 
while PalmTree has a higher recall and doc2vec has a higher precision. 
However, since PalmTree is a transformer-based solution, 
it will have a relatively low runtime efficiency.
Overall, we choose doc2vec as the embedding technique for \sysname.
}

\begin{table}[h]
\setlength{\abovecaptionskip}{0cm}
\setlength{\belowcaptionskip}{0cm}
\caption{\added{Zero-shot evaluation of embedding methods.}}
\label{tab:embedding}
\centering
\begin{tabular}{c|ccc}
\toprule[1pt]
\added{Embedding}       & \added{Precision} & \added{Recall} & \added{F1}     \\ \hline
\added{word2vec}        & \added{72.5\%}    & \added{78.4\%} & \added{75.3\%} \\
\added{Instruction2Vec} & \added{79.8\%}    & \added{82.7\%} & \added{81.2\%} \\
\added{Asm2Vec}         & \added{92.6\%}    & \added{83.3\%} & \added{87.7\%} \\
\added{PalmTree}        & \added{88.2\%}    & \added{90.1\%} & \added{89.1\%} \\
\added{doc2vec}         & \added{93.0\%}    & \added{85.9\%} & \added{89.3\%} \\ 
\bottomrule[1pt]
\end{tabular}
\end{table}

\end{document}